\documentclass[a4paper,11pt]{article}
\usepackage[normalem]{ulem}
\usepackage{amsmath,amssymb,color,comment}
\usepackage{slashed, tensor, bm, physics}
\usepackage{caption}
\usepackage{graphicx}
\usepackage{multirow}
\usepackage{float}
\usepackage[compat=1.1.0]{tikz-feynhand}
\usepackage{jheppub} 

\usepackage[T1]{fontenc} 
\usepackage{booktabs} 
\usepackage{cancel}
\usepackage{listings}
\usepackage{xcolor}

\newcommand{\mbench}{1~\mathrm{MeV}}
\newcommand{\Tpc}{155~\mathrm{MeV}}
\newcommand{\TswWin}{[120,\,200]~\mathrm{MeV}}
\newcommand{\LchiVal}{1.2~\mathrm{GeV}}
\newcommand{\LchiWin}{[1.0,\,1.4]~\mathrm{GeV}}
\newcommand{\pertFloor}{1~\mathrm{GeV}}

\newcommand{\DfFItotal}{1.20\times10^{10}~\mathrm{GeV}}
\newcommand{\DshareHadronic}{79.8\%}
\newcommand{\DsharePiPi}{61.6\%}
\newcommand{\DshareQuarkDecay}{20.1\%}
\newcommand{\DshareBtoD}{19.5\%}
\newcommand{\DshareMesonDecay}{7.1\%}
\newcommand{\DshareQuarkScatt}{0.006\%}
\newcommand{\DbandTsw}{^{+18\%}_{-19\%}}
\newcommand{\DmesShare}{72.7\%}

\newcommand{\UfFItotal}{1.05\times10^{10}~\mathrm{GeV}}
\newcommand{\UshareHadronic}{99.9\%}
\newcommand{\UsharePiPi}{86.8\%}
\newcommand{\UshareQuarkDecay}{0.13\%}
\newcommand{\UbandTsw}{^{+26\%}_{-29\%}}

\newcommand{\fthD}{1.3\times10^{8}~\mathrm{GeV}}
\newcommand{\fthU}{4.0\times10^{7}~\mathrm{GeV}}

\newcommand{\Nscan}{10^{5}}
\newcommand{\NscanBig}{2\times10^{6}}
\newcommand{\bMin}{0.01}
\newcommand{\bMax}{0.99}

\newcommand{\RbenchD}{20.4}
\newcommand{\RminD}{0.29}
\newcommand{\RminDbig}{0.21}
\newcommand{\fFIbenchD}{1.2\times10^{10}~{\rm GeV}}
\newcommand{\fFIshiftD}{4.5}
\newcommand{\NAsixtwoShift}{384}
\newcommand{\SNshiftD}{31}

\newcommand{\RmaxU}{0.269}
\newcommand{\RadvU}{5.44\times10^{-3}}

\newcommand{\fFIshiftU}{3.9}
\newcommand{\CLEOShift}{4.7\times10^{4}}

\newcommand{\survMassLoU}{15.9~{\rm keV}}
\newcommand{\survMassHiU}{100~{\rm keV}}
\newcommand{\survMassU}{[\survMassLoU,\,\survMassHiU]}
\newcommand{\RubinWDM}{62~{\rm keV}}

\title{Anomaly-free axion-like particle in Nelson-Barr models}
\author{Mohammad Aghaie,}
\author{Ryoma Masuda,}
\author{and Ryosuke Sato}

\affiliation{Department of Physics, The University of Osaka, Toyonaka, Osaka 560-0043, Japan}

\emailAdd{aghaie@het.phys.sci.osaka-u.ac.jp}
\emailAdd{masuda@het.phys.sci.osaka-u.ac.jp}
\emailAdd{rsato@het.phys.sci.osaka-u.ac.jp}

\abstract{
  We study Nelson--Barr models with a discrete $Z_N$ symmetry that solve the strong CP problem through spontaneous CP violation, and show that they naturally predict a light axion-like particle (ALP) without introducing any additional ingredients. Unlike the QCD axion, this ALP is anomaly-free: its couplings to photons and gluons are highly suppressed, rendering it naturally long-lived. Instead, it couples to quarks through flavor-violating interactions whose structure is dictated by the CKM matrix. These interactions induce rare meson decays, providing a unique probe of the Nelson--Barr mechanism. We study the cosmological production of the ALP through both freeze-in and misalignment mechanisms. We show that the parameter space in which the observed relic abundance is explained by the freeze-in mechanism is subject to stringent constraints from precision flavor experiments and stellar cooling bounds from SN1987A, leaving only a small viable region that will be comprehensively tested by future structure-formation observations such as the Vera Rubin Observatory and next-generation X-ray missions like Athena, GECCO and THESEUS. In contrast, misalignment production remains a robust and viable mechanism for explaining the observed dark matter abundance over a broad region of parameter space. Our results demonstrate that precision flavor measurements, cosmological observations, and X-ray searches provide complementary probes of this anomaly-free ALP and, consequently, of the Nelson--Barr solution to the strong CP problem.
  }

\begin{document} 
\begin{flushright}
OU-HET-1320
\end{flushright}
\maketitle
\flushbottom

\section{Introduction}

CP violation in flavor-changing processes has been observed in precision measurements of hadron decays and mixings, and the results can be consistently explained by an ${\cal O}(1)$ complex phase in the Cabibbo–Kobayashi–Maskawa (CKM) matrix. On the other hand, CP violation in flavor-conserving observables has not been observed so far. In particular, measurements of the neutron electric dipole moment provide the most stringent upper bound on the effective $\theta$-angle, $|\bar\theta|\lesssim 10^{-10}$~\cite{Abel:2020pzs}. The origin of such a large hierarchy between the CKM phase and $\bar\theta$ is unclear within the Standard Model, and this is known as the strong CP problem~\cite{Jackiw:1976pf, Callan:1976je}.
A particularly elegant solution is provided by the Nelson--Barr mechanism~\cite{Nelson:1983zb,Barr:1984qx,Barr:1984fh}, in which CP is imposed as an exact symmetry of the ultraviolet theory and broken spontaneously. The vacuum expectation values (VEVs) of complex scalar fields, $\langle \Phi_a \rangle$,  generate the CKM phase, while the special structure of the quark mass matrix guarantees a real determinant and hence $\bar\theta=0$ at tree level. In realistic implementations, however, this structure must be protected against radiative corrections and higher-dimensional operators that can regenerate an unacceptably large $\bar\theta$. One attractive solution to this quality problem is realized in gauge-mediated SUSY breaking models \cite{Dine:1993qm, Hiller:2001qg, Evans:2020vil} with a $Z_N$ symmetry \cite{Dine:2015jga, Fujikura:2022sot}.

In this paper, we show that the same $Z_N$ structure required to realize the Nelson--Barr mechanism naturally predicts a light axion-like particle (ALP).
Previously, ALPs in the Nelson--Barr framework were discussed in the minimal Bento--Branco--Parada (BBP) model~\cite{Bento:1991ez} with a $Z_2$ symmetry in ref.~\cite{Dine:2024bxv}, and in extended models with $Z_4 \times Z_{4n}$ symmetries in refs.~\cite{Murai:2024alz, Murai:2024bjy}.
We observe that, in the class of $Z_N$ Nelson--Barr models considered in this paper, an accidental global $U(1)$ symmetry arises for sufficiently large $N$ without introducing any additional fields or symmetries. An explicit breaking effect of $U(1)$ symmetry first appears through higher-dimensional operators.
Since the scalar fields responsible for spontaneous CP violation also break the $Z_N$ and consequently the approximate $U(1)$ symmetry, its pseudo-Nambu--Goldstone boson can be identified with an ALP. The ALP mass is generated only by the small explicit breaking of the accidental symmetry and is therefore naturally suppressed. Consequently, a light ALP is not an additional assumption but rather a generic low-energy prediction of this class of $Z_N$ Nelson--Barr models.

An important feature of this ALP is that the accidental $U(1)$ symmetry is anomaly-free under the Standard Model gauge group~\cite{Nakayama:2014cza,Takahashi:2020bpq,Han:2020dwo,Han:2022iig,Sakurai:2022roq,Aghaie:2024jkj}. Unlike the QCD axion, its couplings to gluons and photons are therefore highly suppressed, rendering it naturally long-lived. Instead, its leading interactions arise through flavor-violating couplings to quarks inherited from the Nelson--Barr sector. These flavor structures simultaneously determine the experimental signatures of the ALP through rare meson decays---such as $K^{+}\to\pi^{+}\phi$ at NA62~\cite{NA62:2025upx} and $D^{+}\to\pi^{+}\phi$ at CLEO~\cite{MartinCamalich:2020dfe, CLEO:2008ffk}---and its production in the early Universe, establishing an intimate connection between flavor physics and cosmology.

We investigate the cosmological evolution of the anomaly-free ALP and study its production through both the freeze-in and misalignment mechanisms. We show that the parameter space in which the observed relic abundance is explained by the freeze-in mechanism is stringently constrained by precision flavor measurements and stellar cooling bounds from SN1987A~\cite{Carenza:2019pxu}, leaving only a small region of parameter space where freeze-in remains viable. This surviving region is expected to be comprehensively tested by upcoming measurements of cosmic structure formation, specifically through warm-dark-matter limits~\cite{DEramo:2020gpr} and future observations by the Vera C. Rubin Observatory~\cite{Baumholzer:2020hvx}, together with current (e.g. Chandra, XMM-Newton, NuSTAR, and INTEGRAL~\cite{Boddy:2022knd}) and future X-ray line searches (e.g. Athena~\cite{Neronov:2015kca}, GECCO~\cite{Coogan:2021rez} and THESEUS which has three different instruments on board, namely XGIS-S, XGIS-X and SXI~\cite{Thorpe-Morgan:2020rwc, Panci:2022wlc}). In contrast, production through the misalignment mechanism remains viable over a broad parameter space and naturally accounts for the observed dark matter abundance. Our results demonstrate that precision flavor experiments, cosmological observations, stellar bounds, and X-ray searches provide complementary probes of this anomaly-free ALP and, consequently, offer a direct experimental window into the Nelson--Barr solution to the strong CP problem.

The outline of this paper is as follows.
In section~\ref{sec:model}, we briefly describe our setup and the emergence of the ALP.
In section~\ref{sec:EFT}, we discuss the low-energy effective theory of the ALP.
In section~\ref{sec:anomalyfreeALP}, we discuss the ALP as a candidate for dark matter and calculate the relic abundance from the misalignment mechanism and the freeze-in mechanism.

\section{\texorpdfstring{$Z_N$}{ZN} Nelson-Barr models}\label{sec:model}
In this section, we provide a brief description of our setup.
Following the minimal Bento-Branco-Parada (BBP) model~\cite{Bento:1991ez}, we introduce one flavor of $SU(2)_L$ singlet vector-like quark that mixes with the Standard Model (SM) quarks. We refer to the model with a down-type (up-type) vector-like quark as the type-D (type-U) model. We impose $Z_N$ symmetry on models to avoid generating nonzero $\bar\theta$ at tree level \cite{Dine:2015jga}.
For later discussion, we parametrize the coupling constants in the Lagrangian following the discussion in ref.~\cite{Cherchiglia:2020kut}.

\subsection{Type-D model}\label{sec:typeD}
In the type-D model, we introduce one flavor of down-type vector-like quarks, $D_L~(3,1,-1/3)$ and $D_R^c~(\bar 3,1,1/3)$, together with complex scalar fields $\Phi_a~(1,1,0)$ $(a = 1, \cdots, N_\Phi)$.
We impose a $Z_N$ symmetry, under which the fields transform as
\begin{align}
    \Phi_a \to e^{2\pi i / N} \Phi_a, \quad
    D_L \to e^{2\pi i / N} D_L, \quad
    D_R^c \to e^{-2\pi i / N} D_R^c, \label{eq:ZN typeD}
\end{align}
while all other SM fields are neutral under $Z_N$.
The minimal BBP model corresponds to the case with $N=2$ and $N_\Phi = 1$. Here we focus on the case with $N\geq 3$, and the corresponding Lagrangian is given by
\begin{align}
    \mathcal{L} =
    -\bar q_L y_u \tilde H u_R
    -\mqty(\bar{q}_{L} & \bar{D}_{L}
    )\mqty(y_{d}H & 0  \\  \kappa_{a} \Phi_{a} & M_D  )\mqty(d_{R}\\ D_{R}) + h.c., \label{eq:Lagrangian typeD}
\end{align}
where $\tilde H = i \sigma^2 H^*$. In this case, $\Phi_a^* \bar D_L d_R$ interaction is prohibited by $Z_N$ symmetry as opposed to the minimal BBP model. Thus, $N_\Phi \geq 2$ is required to obtain a physical CP phase after the $\Phi_a$ fields acquire their VEVs. We impose CP symmetry so that all parameters $y_{u,ij}$, $y_{d,ij}$, $\kappa_{ai}$, and $M_D$ are real.  
When the scalar fields $\Phi_a$ acquire complex VEVs, CP symmetry is spontaneously broken.
After symmetry breaking, the effective Lagrangian becomes
\begin{align}
    \mathcal{L} =
    -\bar q_L y_u \tilde H u_R
    -\mqty(\bar{q}_{L} & \bar{D}_{L}
    )\mqty(y_{d}H & 0  \\  B_d & M_D  )\mqty(d_{R}\\ D_{R}) + h.c.,
\end{align}
where $B_{d,i} \equiv \kappa_{ai} \langle \Phi_a \rangle$.  
Note that $\bar\theta = 0$ because the $Z_N$ symmetry forbids the $\bar q_L D_R H$ term, and both $y_d$ and $M_D$ are real. We assume the scale of spontaneous CP breaking is much higher than the electroweak scale, i.e., there is a hierarchy between mass parameters as $M_D,~B_d \gg M_d \equiv y_d \langle H \rangle$. 

Let us discuss how $y_d$, $B_d$, and $M_D$ are related to the SM quark masses and the CKM matrix.
In the following, we choose a basis in which $y_u$ is diagonal.
We then define the $4\times4$ mass matrix as
\begin{align}
    \mathcal{M}_d=\mqty(M_{d} & 0 \\ B_d & M_D  ),\quad
    M_{d}=y_{d}\ev{H},\quad
    B_d = \kappa_a \ev{\Phi_a}. \label{eq:mass matrix typeD}
\end{align}
We denote mass eigenstates as $\widehat d_{L,R}$ and $\widehat D_{L,R}$, which are related to the original basis $d_{L,R}$ and $D_{L,R}$ by $4\times 4$ unitary matrices $U_L$ and $U_R$:
\begin{align}
    \mqty( d_L \\ D_L ) = U_L^{(D)} \mqty( \widehat d_L \\ \widehat D_L ), \qquad
    \mqty( d_R \\ D_R ) = U_R^{(D)} \mqty( \widehat d_R \\ \widehat D_R ).
\end{align}
The mass matrix $\mathcal{M}$ is then diagonalized as
\begin{align}
    U_{L}^{(D)\dagger}\mathcal{M}_d U_{R}^{(D)}=\mqty(\widehat{M}_{d} &  \\     &   \widehat{M}_D),
\end{align}
where $\widehat{M}_d = {\rm diag}(m_d,m_s,m_b)$. The heavy mass eigenvalue $\widehat{M}_D$ is given by
\begin{align}
    \widehat{M}_D = M_{\rm CP} + {\cal O}(m_b^2 / M_{\rm CP}),
\end{align}
where we defined $M_{\rm CP} \equiv \sqrt{M_D^2 + B_d B_d^\dagger}$.

Assuming $M_D,~B_d \gg M_d$, the unitary matrix $U_{L}^{(D)}$ can be written as
\begin{align}
    U_{L}^{(D)} = \mqty(\vb{1}_{3} & M_d B_d^\dagger / \widehat{M}_D^2 \\  -B_d M_d^T / \widehat{M}_D^2   &  1) \mqty(V_{d_{L}} &  \\     &   1) + {\cal O}(m_b^2/\widehat{M}_D^2).    \label{eq:ULmatrix}
\end{align}
Here $V_{d_L}$ is a unitary matrix satisfying
\begin{align}
    V_{d_L}^\dagger \left( M_{d}M_{d}^{T}-\frac{M_{d} B_d^{\dagger}B_d M_{d}^T}{M_{\rm CP}^2} \right) V_{d_L} = \widehat M_d^2. \label{eq: consistency of SM d-mass}
\end{align}
$V_{d_{L}}$ can be written by using $V_{\rm CKM}$ and two phase parameters $\beta_2$ and $\beta_3$ as
\begin{align}
    V_{d_{L}} = \left(\begin{array}{ccc}
    1 &&\\
    & e^{i\beta_2} & \\
    && e^{i\beta_3}
    \end{array}\right)
    V_{\mathrm{CKM}}.  \label{eq:Vd and CKM}
\end{align}
The unitary matrix $U_R$ is then obtained as
\begin{align}
    U_R^{(D)} &=
    {\cal M}_d^{-1} U_L^{(D)} \left(\begin{array}{cc}
        \widehat M_d & \\
        & \widehat M_D
    \end{array}\right) \nonumber\\ 
    &= \left(\begin{array}{cc}
        M_d^{-1} V_{d_L} \widehat M_d & B_d^\dagger / M_{\rm CP} \\
        -B_d M_d^{-1} V_{d_L} \widehat M_d / M_D & M_D/M_{\rm CP}
    \end{array}\right) + {\cal O}(m_b/M_{\rm CP}). \label{eq:UR typeD}
\end{align}

Next, let us parametrize $M_d$ and $B_d$ such that eq.~\eqref{eq: consistency of SM d-mass} is satisfied. We decompose $B_d$ as
\begin{align}
    B_d = M_{CP} W, \label{eq:B by W}
\end{align}
so that eq.~\eqref{eq: consistency of SM d-mass} can be rewritten as
\begin{align}
    M_{d}(\vb{1}_3-W^{\dagger}W)M_{d}^{T}
    = V_{d_L} \widehat M_d^2 V_{d_L}^\dagger. \label{eq: consistency of SM mass2}
\end{align}
Under the phase redefinition $D_{L,R} \to e^{i\alpha} D_{L,R}$, $W$ transforms as $e^{i\alpha}W$. By choosing an appropriate $\alpha$, one can impose
\begin{equation}
  \Re(W)\cdot\Im(W)=0.
\end{equation}
Furthermore, by an appropriate $SO(3)$ redefinition of $d_{R}$, $W$ can be brought to the form
\begin{align}
    W = ( 0, -ib , a), \quad a,b \in \mathbb{R}. \label{eq:W vector}
\end{align}
Using the definition of $M_{\rm CP}$, eq.~\eqref{eq:B by W}, and eq.~\eqref{eq:W vector}, we obtain
\begin{align}
    M_D = M_{\rm CP} \sqrt{1 - |W|^2} = M_{\rm CP} \sqrt{1 - a^2 - b^2}. \label{eq:MD solution}
\end{align}

In this basis, eq.~\eqref{eq: consistency of SM mass2} becomes
\begin{align}
    M_{d}^{-1} V_{d_L} \widehat M_d^2 V_{d_L}^\dagger (M_d^T)^{-1}
    &= \mqty(1&0&0 \\ 0&1-b^2&-iab\\0&iab&1-a^2).  \label{eq: consistency of SM mass3}
\end{align}
Let us define $X$ and $Y$ as
\begin{align}
    X &\equiv \sqrt{\Re{V_{d_L}\widehat M_d^2 V_{d_L}^\dagger}}, \label{eq:definition X}\\
    Y &\equiv \mathrm{diag}(1, \sqrt{1-b^2} , \sqrt{1-a^2}). \label{eq:definition Y}
\end{align}
Note that $X$ is a real symmetric matrix. Here we choose $X$ such that $\det X > 0$.
Then, the real and imaginary parts of eq.~\eqref{eq: consistency of SM mass3} give
\begin{align}
    M_{d}^{-1} X^2 (M_d^T)^{-1} &
    =Y^2, \label{eq:Re} \\
    M_{d}^{-1} \Im{V_{d_L}\widehat M_d^2 V_{d_L}^\dagger} (M_d^T)^{-1} &= \mqty(0 & 0 & 0 \\ 0 & 0 & -ab\\ 0 & ab & 0). \label{eq:Im}
\end{align} 
A general solution of eq.~\eqref{eq:Re} for $M_d$ is expressed as
\begin{align}
    M_d = X O Y^{-1}, \label{eq:md solution}
\end{align}
where $O$ is a real orthogonal matrix.
Substituting this into eq.~\eqref{eq:Im}, we obtain
\begin{align}
    O^T Z O = \left(\begin{array}{ccc}
        0 & 0 & 0 \\
        0 & 0 & -\mu \\
        0 & \mu & 0
    \end{array}\right), \label{eq:P C mu}
\end{align}
where we defined $Z$ and $\mu$ as
\begin{align}
    Z &\equiv X^{-1}\Im{V_{d_L}\widehat M_d^2 V_{d_L}^\dagger}X^{-1}, \label{eq:definition Z}\\
    \mu &\equiv \frac{ab}{\sqrt{(1-a^2)(1-b^2)}}. \label{eq:a b mu}
\end{align}
Note that $Z$ is a real antisymmetric matrix determined by $m_d$, $m_s$, $m_b$, and $V_{\rm CKM}$. Since $\mu$ can be extracted from $Z$ as
\begin{align}
    \mu = \sqrt{ -\frac{1}{2}{\rm tr}[Z^2] }, \label{eq:mu from C}
\end{align}
$\mu$ is a function of $m_d$, $m_s$, $m_b$, $V_{\rm CKM}$, $\beta_2$, and $\beta_3$.
Consequently, once $b$, $\beta_2$, and $\beta_3$ is fixed, $a$ is determined.
The possible range of $\mu$ is $0<\mu<1$, and for given $\mu$, the possible range of $a$ and $b$ is $0<a<1$ and $0<b<1$. For details, see appendix \ref{app:mu a b}.

Now let us discuss how to determine $O$ so that eq.~\eqref{eq:P C mu} is satisfied. For the matrix $Z$, we can find a unit eigenvector corresponding to the zero eigenvalue as
\begin{align}
    \vec e_1 &= \frac{1}{\sqrt{(Z_{23})^2+(Z_{13})^2+(Z_{12})^2}}
    \mqty(-Z_{23}\\Z_{13}\\-Z_{12}). \label{eq:e1}
\end{align}
Next, we define two orthonormal vectors $e_2$ and $e_3$ that are orthogonal to $e_1$:
\begin{align}
    \vec e_2 &= \frac{1}{\sqrt{(Z_{12}Z_{23})^2+(Z_{12}Z_{13})^2+((Z_{13})^2+(Z_{23})^2)^2}}
    \mqty(-Z_{12}Z_{23}\\Z_{12}Z_{13}\\(Z_{13})^2+(Z_{23})^2 ), \label{eq:e2}\\
    \vec e_3 &= \frac{1}{\sqrt{(Z_{13})^2+(Z_{23})^2}}
    \mqty(Z_{13}\\Z_{23}\\0). \label{eq:e3}
\end{align}
It can be easily verified that $Z \vec e_2 = \mu \vec e_3$ and $Z \vec e_3 = -\mu \vec e_2$. 
A general solution for $O$ that satisfies eq.~\eqref{eq:P C mu} can then be parameterized by an additional angle $\gamma$ as
\begin{align}
    O &= \mqty( \vec e_1 & \vec e_2 & \vec e_3 ) \mqty(1 & & \\ & \cos{\gamma} & \sin{\gamma}\\ & -\sin{\gamma} & \cos{\gamma}). \label{eq:solution O}
\end{align}
Note that $\det O = 1$ and then $\det \mathcal{M}_d = M_D \det M_d$ is real positive.

To summarize, once the parameters 
$M_{\rm CP}$, $\gamma$, $b$, $\beta_2$, and $\beta_3$ are specified, $M_d$, $B$, and $M_D$ in eq.~\eqref{eq:mass matrix typeD}, which reproduce the observed quark masses and $V_{\rm CKM}$, can be obtained through the following procedure:
\begin{enumerate}
    \item Calculate $X$ from eq.~\eqref{eq:definition X}.
    \item Calculate $Z$ from $X$ and eq.~\eqref{eq:definition Z}, and then determine $\mu$ from eq.~\eqref{eq:mu from C}.
    \item Determine $a$ from eq.\eqref{eq:a b mu}, and then obtain $W$ and $Y$  from eq.~\eqref{eq:W vector} and eq.~\eqref{eq:definition Y}.
    \item Construct $O$ from eqs.~(\ref{eq:e1}, \ref{eq:e2}, \ref{eq:e3}, \ref{eq:solution O}).
    \item Finally, calculate $M_d$ from $X$, $Y$, $O$, and eq.~\eqref{eq:md solution}, $B$ from $W$ and eq.~\eqref{eq:B by W}, $M_D$ from $W$ and eq.~\eqref{eq:MD solution}.
\end{enumerate}

\subsection{Type-U model}\label{sec:typeU}
Similarly, in the type-U model, we introduce one flavor of up-type vector-like quarks, $U_L~(3,1,2/3)$ and $U_R^c~(\bar 3,1,-2/3)$, together with complex scalar fields $\Phi_a~(1,1,0)$ $(a = 1, \cdots, N_\Phi)$. 
We again assume a $Z_N$ symmetry under which the fields transform as
\begin{align}
    \Phi_a \to e^{2\pi i / N} \Phi_a, \quad
    U_L \to e^{2\pi i / N} U_L, \quad
    U_R^c \to e^{-2\pi i / N} U_R^c, \label{eq:ZN typeU}
\end{align}
while all other SM fields are neutral under $Z_N$.
Same as the type-D model, we focus on the case with $N\geq 3$ and $N_\Phi \geq 2$, and the Lagrangian is then given by
\begin{align}
    \mathcal{L} =
    -\mqty(\bar{q}_{L} & \bar{U}_{L}
    )\mqty(y_{u}\tilde H & 0  \\  \kappa_a \Phi_a & M_U  )\mqty(u_{R}\\ U_{R})
    -\bar q_L y_d H d_R
+ h.c., \label{eq:Lagrangian typeU}
\end{align}
where $\tilde H = i \sigma^2 H^*$.  
Again, we impose CP symmetry, such that all parameters $y_{u,ij}$, $y_{d,ij}$, $\kappa_{ai}$, and $M_U$ are real.  
CP symmetry is then spontaneously broken when the scalar fields $\Phi_a$ acquire complex vacuum expectation values (VEVs).
After symmetry breaking, the effective Lagrangian becomes
\begin{align}
    \mathcal{L} =
    -\mqty(\bar{q}_{L} & \bar{U}_{L}
    )\mqty(y_{u}\tilde H & 0  \\  B_u & M_U  )\mqty(u_{R}\\ U_{R})
    -\bar q_L y_d H d_R
+ h.c.,
\end{align}
where $B_{u,i} \equiv \kappa_{ai} \langle \Phi_a \rangle$.  
Note that $\bar\theta = 0$ because the $Z_N$ symmetry forbids the term $q_L U_R H$.
We assume the scale of spontaneous CP breaking is much higher than the electroweak scale, i.e., there is a hierarchy between mass parameters as $M_U,~B_u \gg M_u \equiv y_u \langle \tilde H \rangle$. 

Let us discuss how $y_u$, $B_u$, and $M_U$ are related to the SM quark masses and the CKM matrix.
This procedure is almost parallel to the procedure in the type-D model.
In the following, we choose a basis in which $y_d$ is diagonal.
We  define the $4\times4$ mass matrix ${\cal M}_u$ as
\begin{align}
    \mathcal{M}_u=\mqty(M_{u} & 0 \\ B_u & M_U  ),\quad
    M_{u} = y_{u}\tilde{\ev{H}},\quad
    B_u = \kappa_a \ev{\Phi_a}. \label{eq:mass matrix typeU}
\end{align}
The mass matrix $\mathcal{M}_u$ is then diagonalized as
\begin{align}
    U_{L}^{(U)\dagger}\mathcal{M}_u U_{R}^{(U)}=\mqty(\widehat{M}_{u} &  \\     &   \widehat{M}_U),
\end{align}
where $\widehat{M}_{u} = \mathrm{diag}(m_{u}, m_{c}, m_{t})$.
The heavy mass eigenvalue $\widehat M_U$ is given by
\begin{align}
    \widehat M_U = M_{\rm CP} + {\cal O}(m_t^2 / M_{\rm CP}),
\end{align}
where we defined $M_{\rm CP} \equiv \sqrt{M_U^2 + B_u B_u^\dagger}$. 

Assuming $M_U,~ B_u \gg M_u$, the unitary matrix $U_{L}^{(U)}$ can be written as
\begin{align}
    U_{L}^{(U)} = \mqty(\vb{1}_{3} & M_u B_u^\dagger / M_{\rm CP}^2 \\  -B_u M_u^T / M_{\rm CP}^2   &  1) \mqty(V_{u_{L}} &  \\     &   1) + {\cal O}(m_t^2/M_{\rm CP}^2).    
\end{align}
Here $V_{u_L}$ is a unitary matrix satisfying
\begin{align}
    V_{u_L}^\dagger \left( M_{u}M_{u}^{T}-\frac{M_{u} B_u^{\dagger} B_u M_{u}^T}{M_{\rm CP}^2} \right) V_{u_L} = \widehat M_u^2. \label{eq: consistency of SM u-mass}
\end{align}
$V_{u_{L}}$ can be written by using $V_{\rm CKM}$ and two phase parameters $\beta_2$ and $\beta_3$ as
\begin{align}
    V_{u_{L}}^\dagger = V_{\mathrm{CKM}}
    \left(\begin{array}{ccc}
        1 &&\\
        & e^{i\beta_2} & \\
        && e^{i\beta_3}
    \end{array}\right). \label{eq:Vu and CKM}
\end{align}
Thus, comparing type-D and type-U model, although eq.~\eqref{eq:Vd and CKM} and eq.~\eqref{eq:Vu and CKM} have a slight difference, the remaining part is completely same as the procedure in type-D with replacing $y_d$, $B_d$, $M_D$, and $V_{d_L}$ to $y_u$, $B_u$, $M_U$, and $V_{u_L}$.
The unitary matrix $U_R^{(U)}$ is then obtained as
\begin{align}
    U_R^{(U)} &=
    {\cal M}_u^{-1} U_L^{(U)} \left(\begin{array}{cc}
        \widehat M_u & \\
        & \widehat M_U
    \end{array}\right) \nonumber\\ 
    &= \left(\begin{array}{cc}
        M_u^{-1} V_{u_L} \widehat M_u & B_u^\dagger / M_{\rm CP} \\
        -B_u M_u^{-1} V_{u_L} \widehat M_u /M_U & M_U/M_{\rm CP}
    \end{array}\right) + {\cal O}(m_t/M_{\rm CP}). \label{eq:UR typeU}
\end{align}

As we have discussed in section \ref{sec:typeD} for the type-D model, we can construct $M_u$, $B_u$, and $M_U$ from the SM up-type quark masses, $V_{\rm CKM}$, and four free parameters $\beta_2$, $\beta_3$, $b$, and $\gamma$. The construction is the same as the discussion after eq.~\eqref{eq:B by W} with replacements $V_{d_L} \to V_{u_L}$, $M_d \to M_u$, and $M_D \to M_U$.
  
\subsection{Quality problem in the minimal \texorpdfstring{$Z_N$}{ZN} Nelson-Barr model}
Although the structure of the mass matrices given in eq.~\eqref{eq:mass matrix typeD} and eq.~\eqref{eq:mass matrix typeU} guarantees $\bar\theta = 0$ at tree-level, nonzero $\bar\theta$ can be induced from radiative corrections and higher-dimensional effective interactions. This problem has been known as the quality problem and discussed in the literature. For details, see, e.g., refs.~\cite{Dine:2015jga, Perez:2020dbw, Asadi:2022vys}.
Here we briefly comment on the quality problem in the minimal $Z_N$ Nelson--Barr models.

\subsubsection*{One-loop diagram with Higgs portal coupling}
The following Higgs portal coupling of $\Phi$ is also allowed under $Z_N$ symmetry:
\begin{align}
    {\cal L} \ni \lambda_{ab} \Phi_a \Phi_b^* |H|^2. \label{eq:phi higgs}
\end{align}
This coupling induces a one-loop correction to the Yukawa coupling as \cite{Bento:1991ez,Dine:2015jga}
\begin{align}
    \delta y_{ij} \sim \frac{\lambda_{ab} y_{ik} \kappa^*_{ck} \kappa_{bj} \langle \Phi_c^* \rangle \langle \Phi_a \rangle }{16\pi^2 M_\Phi^2}.
\end{align}
Then, we obtain $\delta \bar\theta$:
\begin{align}
    \delta \theta = {\rm Im} \qty[{\rm tr} (y^{-1} \delta y)] \sim \frac{\lambda \kappa^2 \langle \Phi \rangle^2 }{16\pi^2 M_\Phi^2}.
\end{align}
By using $M_Q \sim \widehat M_Q \sim \kappa \langle \Phi \rangle$, we obtain the upper bound on $\lambda_{ab}$ as
\begin{align}
    \lambda_{ab} \lesssim 10^{-8} \times \frac{M_\Phi^2}{\widehat M_Q^2}. \label{eq:quality higgsportal}
\end{align}
Note that a natural value of $\lambda_{ab}$ is of the order of $\langle H\rangle^2 / \langle\Phi\rangle^2$ to realize $\langle H \rangle \ll \langle\Phi\rangle$ \cite{Bento:1991ez}. Thus, this correction decouples by assuming $\langle H \rangle \ll \langle\Phi\rangle$.

\subsubsection*{Two-loop `dead duck' diagram}
Non-decoupling effect on $\delta\bar\theta$ comes from the self-interaction of $\Phi$:
\begin{align}
    {\cal L} \ni \gamma_{abcd} \Phi_a \Phi_b \Phi_c^* \Phi_d^*. \label{eq:phi self}
\end{align}
Two-loop `dead duck' diagram \cite{Nelson:1983zb} gives a correction to the mass of vector-like quark:
\begin{align}
    \delta M_Q \sim M_Q \frac{g_s^2 \gamma}{(16\pi^2)^2} \frac{\kappa^2 \langle\Phi\rangle^2}{{\rm Max}[M_Q^2,M_\Phi^2]}.
\end{align}
Thus, we obtain $\delta \bar\theta$ as
\begin{align}
    \delta\bar\theta = \frac{{\rm Im}[\delta M_Q]}{M_Q} \sim \frac{g_s^2 \gamma}{(16\pi^2)^2} \frac{\kappa^2 \langle\Phi\rangle^2}{{\rm Max}[M_Q^2,M_\Phi^2]}.
\end{align}
By using $M_Q \sim \widehat M_Q \sim \kappa \langle \Phi \rangle$, we obtain the upper bound on $\gamma_{abcd}$ as
\begin{align}
    \gamma_{abcd} \lesssim 10^{-6} \times {\rm Max}\left[ \frac{M_\Phi^2}{M_Q^2}, 1 \right]. \label{eq:quality selfcoupling}
\end{align}
This effect does not decouple in the limit of a large Nelson--Barr scale.
In general, $\gamma \sim {\cal O}(1)$ and this requires $M_Q / M_\Phi \lesssim 10^{-3}$ or equivalently $\kappa \lesssim 10^{-3}$ \cite{Perez:2020dbw}.

\subsubsection*{Dimension-5 operators}
In addition to radiative corrections, higher-dimensional effective interactions suppressed by the cutoff scale of the theory can also induce $\delta\bar\theta$.
The following dimension-5 effective interactions are allowed by the $Z_N$ symmetry:
\begin{align}
    {\cal L} \sim \frac{c_{ab}}{\Lambda} \Phi_a \Phi_b^* Q_L Q_R^c + \frac{c'_{ai}}{\Lambda} \Phi^*_a \bar q_L H Q_R,\label{eq:dim5 op}
\end{align}
where $Q = U~{\rm or}~D$ and $\Lambda$ is a cutoff scale of the theory.
These terms break the structure of the quark mass matrix given in eq.~\eqref{eq:mass matrix typeD} and eq.~\eqref{eq:mass matrix typeU}.
As a result, we obtain the correction to $\bar\theta$ as \cite{Asadi:2022vys}
\begin{align}
    \delta\bar\theta \sim \frac{c_{ab} \langle \Phi_a \rangle \langle \Phi_b^* \rangle}{M_Q \Lambda} + \frac{B_{i} y^{-1}_{ij} c'_{aj} \langle \Phi_a \rangle}{M_Q \Lambda}.
\end{align}
By using $M_Q \sim \kappa \langle\Phi\rangle = B$, we obtain $\delta\bar\theta$ as \begin{align}
    \delta\bar\theta \sim \frac{\langle \Phi\rangle}{\Lambda}  \times \frac{1}{{\rm min}[y,\kappa]}.
\end{align}
By assuming $\Lambda \lesssim M_{\rm pl}$ and $\delta\bar\theta \lesssim 10^{-10}$, we obtain \begin{align}
    \langle \Phi \rangle \lesssim 10^8~{\rm GeV} \times {\rm min}[y,\kappa]. \label{eq:quality dim5}
\end{align}
This gives a severe upper bound on the Nelson--Barr energy scale. See also refs.~\cite{Dine:2015jga, Perez:2020dbw}.

\subsubsection*{Solutions to the quality problem}
To solve the strong CP problem by the Nelson--Barr mechanism, we need to suppress the dangerous corrections $\delta\bar\theta$ described so far. Let us briefly summarize the existing proposal to address this problem.

In general, eq.~\eqref{eq:quality selfcoupling} and eq.~\eqref{eq:quality dim5} give severe constraints on Nelson--Barr models \cite{Perez:2020dbw, Asadi:2022vys}.
First, the constraint in eq.~\eqref{eq:quality dim5} with $y_{u/d} \sim 10^{-5}$ requires $\langle \Phi \rangle \lesssim 10^3~{\rm GeV}$.
In addition, $\gamma_{abcd} \sim {\cal O}(1)$ with eq.~\eqref{eq:quality selfcoupling} requires $\kappa \lesssim 10^{-3}$ and this leads to $M_Q \lesssim 1~{\rm GeV}$! Apparently, this destroys the unitarity of the CKM matrix, and such a light exotic colored particle has not been observed at collider experiments.
Viable models can be constructed, for example, by assuming an extra symmetry \cite{Perez:2020dbw, Asadi:2022vys, Perez:2023zin, Murai:2024alz}, a five-dimensional model \cite{Girmohanta:2022giy},
and CP breaking mediation by higher-dimensional operators \cite{Valenti:2021xjp, Csaki:2025ikr}.

Supersymmetry (SUSY) is one of the simplest solutions to solve the quality problem. Since the interaction terms in eqs.~\eqref{eq:phi higgs}, \eqref{eq:phi self}, and \eqref{eq:dim5 op} have a non-holomorphic dependence on $\Phi_a$, their structure can be naturally controlled~\cite{Dine:1993qm}.
Since additional sources of $\bar\theta$, such as the gluino mass, can appear in SUSY models, gauge mediation is one of the favored SUSY-breaking mediation mechanisms to avoid this issue \cite{Hiller:2001qg, Dine:2015jga, Evans:2020vil, Fujikura:2022sot}.

In the next section, we will show that an axion-like particle naturally arises in $Z_N$ Nelson--Barr models with relatively large $N$, and its properties and phenomenology are insensitive to the details of the model. Thus, in the remainder of this paper, we will just assume that some extension of the current model solves the quality problem and we neglect the constraints on the minimal model given in eqs.~\eqref{eq:quality higgsportal}, \eqref{eq:quality selfcoupling}, and \eqref{eq:quality dim5}.

\section{ALP and its Effective Lagrangian}\label{sec:EFT}
In this section, we describe how an axion-like particle (ALP) arises in both the type-D and type-U models which are introduced in the previous section, and derive an effective Lagrangian of ALP.

\subsection{ALP}\label{sec:ALP}
Let us now discuss the following $U(1)$ transformation:
\begin{align}
    \Phi_a \to e^{i\alpha} \Phi_a, \quad
    Q_L \to e^{i\alpha} Q_L, \quad
    Q_R^c \to e^{-i\alpha} Q_R^c,
\end{align}
where $Q = D~(U)$ for the type-D (type-U) model.  
The $Z_N$ symmetries described in eqs.~\eqref{eq:ZN typeD} and \eqref{eq:ZN typeU} can be understood as subgroups of this $U(1)$ symmetry.  
For $N \ge 5$, this $U(1)$ symmetry arises accidentally, since no renormalizable interactions among the SM fields, vector-like quarks, and $\Phi_a$ can break it.  
The $U(1)$ symmetry is explicitly broken down to $Z_N$ by dimension-$N$ effective operators:
\begin{align}
    {\cal L} \sim \frac{1}{N! \Lambda^{N-4}}\, \Phi_a^N + \cdots, \label{eq:U(1)breaking_term}
\end{align}
where $\Lambda$ denotes the cutoff scale of the effective theory.  

For relatively large $N$, the explicit breaking of $U(1)$ given in eq.~\eqref{eq:U(1)breaking_term} is suppressed by a power of $\Lambda$. Thus, we can understand that the $U(1)$ symmetry is an approximate symmetry that is spontaneously broken by the VEV of $\Phi_a$'s. This means there exists a pseudo Nambu--Goldstone boson $\phi$ associated with this $U(1)$ symmetry.
By using this $\phi$, $\Phi_a$ are parametrized as
\begin{align}
    \Phi_a = \langle \Phi_a \rangle \exp\left( i \phi / f \right), \qquad
    f \equiv \sqrt{ 2 \sum_a |\langle \Phi_a \rangle|^2 }. \label{eq:def ALP}
\end{align}
The normalization of $f$ is chosen such that the field $\phi$ has a canonical kinetic term.  
The resulting potential for $\phi$ is then given by
\begin{align}
    V \sim -\frac{f^N}{2^{N/2} \,  N! \,  \Lambda^{N-4}} \cos\!\left( \frac{N\phi}{f} \right),
\end{align}
and the corresponding mass is
\begin{align}
    m_\phi^2 \sim \frac{N^2 f^{N-2}}{2^{N/2} \, N! \,\Lambda^{N-4}}. \label{eq:ALPmass}
\end{align}
Thus, for sufficiently large $N$ and $\Lambda$, the mass of $\phi$ becomes naturally small.
In the following of this paper, we call $\phi$ an axion-like particle (ALP).

\subsection{The effective field theory below \texorpdfstring{$f$}{f} }
In the current models, the vector-like quark and particles in the $\Phi$ sector, except for the ALP $\phi$, have masses of the order of $f$. 
Thus, we can integrate out those heavy particles and construct the effective field theory for the SM particles and $\phi$ with the cutoff scale $\sim f$.

\subsubsection{Type-D model}
First, we discuss the EFT for the type-D model.
The ALP field $\phi$ originates from the phase of the complex scalar fields $\Phi_a$ as eq.~\eqref{eq:def ALP}, and this can be absorbed by $U(1)$ transformation as
\begin{align}
    D_{L,R} \to \exp\left( \frac{i\phi}{f} \right) D_{L,R}, 
\end{align}
then we obtain the following derivative coupling between $a$ and $D$ as
\begin{align}
    {\cal L} = \frac{1}{f} (\partial_\mu \phi) ( \bar{D}_L \gamma^\mu D_L + \bar{D}_R \gamma^\mu D_R ). \label{eq:ALP_quark_coupling_typeD}
\end{align}
Transforming to the quark mass basis, the derivative interaction becomes
\begin{align}
    {\cal L} = \frac{1}{f}(\partial_\mu\phi) \left( \bar{\widehat d}_L\gamma^\mu U_L^{(D)\dagger} Q U_L^{(D)} \widehat d_L + \bar{\widehat d}_R\gamma^\mu U_R^{(D)\dagger} Q U_R^{(D)} \widehat d_R \right),
\end{align}
where $Q = \mathrm{diag}(0,0,0,1)$ projects onto the vector-like quark. Since the mixing between the left-handed Standard Model quarks and the vector-like quark is suppressed by $\mathcal{O}(m_d/M_D)$, the resulting left-handed ALP couplings are suppressed by $\mathcal{O}(m_d^2/M_D^2)$ and can be neglected. In contrast, the right-handed mixing is unsuppressed by the light quark masses and generates the leading flavor-violating interactions among the light quarks,
\begin{align}
    {\cal L } &\ni \frac{C^{(D)}_{ij}}{f} (\partial_{\mu}\phi)  \overline{{\widehat{d}}}_{R{i}} \gamma^{\mu}\widehat{d}_{Rj}, \label{eq:ALP quark coupling type-D}
\end{align}
where $C^{(D)}_{ij}$ is written by using $U_R^{(D)}$ in eq.~\eqref{eq:UR typeD} as
\begin{align}
    C^{(D)}_{ij}
    & = U_{R,4i}^{(D)*} U_{R,4j}^{(D)}
    \simeq
    \frac{1}{M^2_{D}} \left( \widehat{M}_{d}V^{\dagger}_{d_L}(M^{-1}_{d})^{T}B^{\dagger}B M_d^{-1} V_{d_L}\widehat{M_{d}} \right)_{ij}. \label{eq:CDij}
\end{align}

The interaction terms between the ALP and light mesons can be read off from the chiral Lagrangian :
\begin{equation}
\begin{split}
  \mathcal{L}_{\rm eff.}&=\frac{f^2_{\pi}}{4}\mathrm{Tr}\qty[D^{\mu}\Sigma(D_{\mu}\Sigma)^{\dagger}] +\frac{f^2_{\pi}}{2}B_{0}\mathrm{Tr}\qty[\mathcal{M}_{q}\Sigma^{\dagger} + h.c.]-\frac{1}{2}M^2_{0}\eta^2_{0}
  + \frac{1}{2}(\partial^{\mu}\phi)(\partial_{\mu}\phi), \label{eq: Chiral Lag}
\end{split}
\end{equation}
where ${\cal M}_q = (m_u, m_d, m_s)$ and $f_\pi = 93~{\rm MeV}$.
For details of the matching procedure between the EFT and the UV Lagrangian, see the appendix \ref{app:meson_eft}.
The meson field $\Sigma$ is defined as
\begin{align}
  \Sigma(x) = \exp\qty[i\frac{\sqrt{2}}{f_{\pi}}\Pi(x)]; \qquad
  \Pi(x) = \mqty(\frac{\pi^0}{\sqrt{2}}+ \frac{\eta^8}{\sqrt{6}}+\frac{\eta^0}{\sqrt{3}}& \pi^{+} & K^{+} \\  \pi^{-}  & -\frac{\pi^{0}}{\sqrt{2}}+\frac{\eta^8}{\sqrt{6}}+\frac{\eta^0}{\sqrt{3}} & K^{0} \\ 
  K^{-} & \bar{K}^0 & -\frac{\eta^8}{\sqrt{3/2}}+\frac{\eta^0}{\sqrt{3}}).
\end{align}
The effect of the ALP quark coupling in eq.~\eqref{eq:ALP quark coupling type-D} is written in the form of the covariant derivative as \cite{Gasser:1984gg, Bauer:2021wjo, Aghaie:2024jkj}
\begin{equation}
    D_{\mu}\Sigma = \partial_{\mu}\Sigma + \frac{i}{f} (\partial_{\mu}\phi)\Sigma C'^{(D)}.
\end{equation}
$C'^{(D)}$ is the matrix obtained by transforming $C^{(D)}$ from the $(d,s,b)$ basis to the $(u,d,s)$ basis as 
\begin{equation}
  C'^{(D)} = \mqty( 0 & 0 & 0 \\ 0 & C^{(D)}_{dd} & C^{(D)}_{ds} \\ 0 & C^{(D)}_{sd} & C^{(D)}_{ss}).
\end{equation}
The coupling terms of ALP and light mesons at the leading order are
\begin{align}
  \mathcal{L}
  &\ni -\frac{if^2_{\pi}}{4f} (\partial_{\mu}\phi) {\rm tr}[(\partial^{\mu}\Sigma) C'^{(D)} \Sigma^{\dagger}] + h.c. \nonumber\\
  &\simeq \frac{f_{\pi}}{\sqrt{2}f}  \Bigg\{ -\frac{C^{(D)}_{dd}}{\sqrt{2}}(\partial_{\mu}\phi) (\partial^{\mu}\pi^0) + \qty(\frac{C^{(D)}_{dd}}{\sqrt{6}} -\frac{C^{(D)}_{ss}}{\sqrt{3/2}} )(\partial_{\mu}\phi) (\partial^{\mu}\eta^8) + \frac{C^{(D)}_{dd}+C^{(D)}_{ss}}{\sqrt{3}}(\partial_{\mu}\phi) (\partial^{\mu}\eta^0) \nonumber\\
  &\hspace{45pt}+ C^{(D)}_{sd}(\partial_{\mu}\phi) (\partial^{\mu}K^0) + C^{(D)}_{ds}(\partial_{\mu}\phi) (\partial^{\mu}\bar{K}^0) \Bigg\}. \label{eq:ALP meson interaction}
\end{align}

The flavor-diagonal coupling between the ALP and the SM quarks induces the effective coupling between the ALP and the photon.
$C_{bb}^{(D)}$ coupling contributes to $\phi\gamma\gamma$ effective interaction via one-loop diagram with bottom quark loop. For $C_{dd}^{(D)}$ and $C_{ss}^{(D)}$, the ALP-meson mixing via eq.~\eqref{eq:ALP meson interaction} contributes to the effective interaction. Then we obtain
\begin{align}
    {\cal L} \ni -\frac{1}{4} g_{\phi\gamma\gamma} \phi F\tilde F + \frac{\alpha_{\rm em}}{144\pi f} \frac{C^{(D)}_{bb}}{m^2_{b}} \qty((\partial^2 \phi)F_{\mu\nu} \tilde{F}^{\mu\nu} + 2\phi F_{\mu\nu} \partial^2 \tilde{F}^{\mu\nu}).
\end{align}
The last term represents the contribution from the bottom quark \cite{Nakayama:2014cza}.
$g_{\phi\gamma\gamma}$ represents the contribution from the ALP-meson mixing and it is written as
\begin{align}
    g_{\phi\gamma\gamma} &= 
    \frac{\alpha_{\mathrm{em}} m_\phi^2}{\pi f_{\pi}} \Biggl[
    \frac{\xi^{(D)}_{\phi\pi^0}}{m_{\pi^0}^2} 
    + \frac{1}{m_{\eta}^2} \left( \xi^{(D)}_{\phi\eta^8} \cos\theta_\eta - \xi^{(D)}_{\phi\eta^0} \sin\theta_\eta \right) \left( \frac{\cos\theta_\eta}{\sqrt{3}} - \frac{4\sin\theta_\eta}{\sqrt{6}} \right) \nonumber\\
&\qquad\qquad\qquad    + \frac{1}{m_{\eta'}^2} \left( \xi^{(D)}_{\phi\eta^8} \sin\theta_\eta + \xi^{(D)}_{\phi\eta^0} \cos\theta_\eta \right) \left( \frac{\sin\theta_\eta}{\sqrt{3}} + \frac{4\cos\theta_\eta}{\sqrt{6}} \right)
    \Biggr], \label{eq:ALPphoton_typeD}
\end{align}
and $\xi^{(D)}_{a\pi^0}$, $\xi^{(D)}_{a\eta^8}$, and $\xi^{(D)}_{a\eta^0}$ are the coefficient of kinetic mixing terms defined as
\begin{align}
    \xi_{\phi\pi^0}^{(D)} = -\frac{C^{(D)}_{dd}}{2}\frac{f_{\pi}}{f},\quad
    \xi_{\phi\eta^8}^{(D)} =  \frac{1}{2\sqrt{3}}\qty(C^{(D)}_{dd} -2\,C^{(D)}_{ss} )\frac{f_{\pi}}{f},\quad
    \xi_{\phi \eta^0}^{(D)} = \frac{1}{\sqrt{6}} \qty(C^{(D)}_{dd}+C^{(D)}_{ss})\frac{f_{\pi}}{f}.
\end{align}

\subsubsection{Type-U model}
Next, we discuss the EFT for the type-U model.
In the case of the type-U model, $\phi$ couples with the SM quarks as
\begin{align}
    {\cal L } &\ni \frac{C^{(U)}_{ij}}{f} (\partial_{\mu}\phi)  \overline{{\widehat{u}}}_{R{i}} \gamma^{\mu}\widehat{u}_{Rj}, \label{eq:ALP quark coupling type-U}
\end{align}
where $C^{(U)}_{ij}$ is written by using $U_R^{(U)}$ in eq.~\eqref{eq:UR typeU} as
\begin{align}
    C^{(U)}_{ij}
    & = U_{R,4i}^{(U)*} U_{R,4j}^{(U)}
    \simeq
    \frac{1}{M^2_{U}} \left( \widehat{M}_{u} V_{u_L}^\dagger (M^{-1}_{u})^{T}B^{\dagger}B M_u^{-1} V_{u_L} \widehat{M_{u}} \right)_{ij}. \label{eq:CUij}
\end{align}
The coupling between the ALP and the light mesons at the leading order is written as
\begin{align}
  \mathcal{L}
  &\ni -\frac{if^2_{\pi}}{4f} (\partial_{\mu}\phi) {\rm tr}[(\partial^{\mu}\Sigma) C'^{(U)} \Sigma^{\dagger}] + h.c. \nonumber\\
  &\simeq \frac{f_{\pi}}{\sqrt{2}f}  \Bigg\{ \frac{C^{(U)}_{uu}}{\sqrt{2}}(\partial_{\mu}\phi) (\partial^{\mu}\pi^0) + \frac{C^{(U)}_{uu}}{\sqrt{6}} (\partial_{\mu}\phi) (\partial^{\mu}\eta^8) + \frac{C^{(U)}_{uu}}{\sqrt{3}}(\partial_{\mu}\phi) (\partial^{\mu}\eta^0) \nonumber\ \Bigg\}. \label{eq:ALP meson interaction U-type}
\end{align}

Similar to the type-D model, the flavor-diagonal coupling between the ALP and the SM induces the effective coupling between the ALP and photon as
\begin{align}
    {\cal L} \ni -\frac{1}{4} g_{\phi\gamma\gamma} \phi F\tilde F + \frac{\alpha_{\rm em}}{36\pi f} \left(\frac{C^{(U)}_{cc}}{m^2_{c}} + \frac{C^{(U)}_{tt}}{m^2_{t}} \right) \qty((\partial^2 \phi)F_{\mu\nu} \tilde{F}^{\mu\nu} + 2\phi F_{\mu\nu} \partial^2 \tilde{F}^{\mu\nu} ).
\end{align}
The last term represents the contribution from the charm and top  \cite{Nakayama:2014cza}. $g_{\phi\gamma\gamma}$ represents the contribution from the ALP-meson mixing and it is written as
\begin{align}
    g_{\phi\gamma\gamma} &= 
    \frac{\alpha_{\mathrm{em}} m_\phi^2}{\pi f_{\pi}} \Biggl[
    \frac{\xi^{(U)}_{\phi\pi^0}}{m_{\pi^0}^2} 
    + \frac{1}{m_{\eta}^2} \left( \xi^{(U)}_{\phi\eta^8} \cos\theta_\eta - \xi^{(U)}_{\phi\eta^0} \sin\theta_\eta \right) \left( \frac{\cos\theta_\eta}{\sqrt{3}} - \frac{4\sin\theta_\eta}{\sqrt{6}} \right) \nonumber\\
&\qquad\qquad\qquad    + \frac{1}{m_{\eta'}^2} \left( \xi^{(U)}_{\phi\eta^8} \sin\theta_\eta + \xi^{(U)}_{\phi\eta^0} \cos\theta_\eta \right) \left( \frac{\sin\theta_\eta}{\sqrt{3}} + \frac{4\cos\theta_\eta}{\sqrt{6}} \right)
    \Biggr], \label{eq:ALPphoton_typeU}
\end{align}
and $\xi^{(U)}_{a\pi^0}$, $\xi^{(U)}_{a\eta^8}$, and $\xi^{(U)}_{a\eta^0}$ are the coefficient of kinetic mixing terms defined as
\begin{align}
    \xi_{\phi\pi^0}^{(U)} = \frac{C^{(U)}_{uu}}{2}\frac{f_{\pi}}{f},\qquad
    \xi_{\phi\eta^8}^{(U)} =  \frac{C^{(U)}_{uu}}{2\sqrt{3}} \frac{f_{\pi}}{f},\qquad
    \xi_{\phi \eta^0}^{(U)} =  \frac{C^{(U)}_{uu}}{\sqrt{6}}\frac{f_{\pi}}{f}.
\end{align}

\section{Anomaly-free ALP dark matter}\label{sec:anomalyfreeALP}
In this section, we discuss the anomaly-free ALP $\phi$ as a dark matter candidate. We calculate its relic abundance from both the misalignment and the freeze-in mechanism, and show the current constraints and future prospects in the parameter space.

\subsection{Relic abundance of ALP dark matter}
\label{sec:relic}
The current models possess $Z_N$ symmetry, which is spontaneously broken in the vacuum. Thus, to avoid the domain wall problem, we assume that the reheating temperature $T_{\rm RH}$ is below the Nelson--Barr scale $f$. We evaluate the relic abundance of ALP in the current universe based on the misalignment mechanism~\cite{Preskill:1982cy, Abbott:1982af, Dine:1982ah} and the freeze-in mechanism~\cite{Hall:2009bx}.

\subsubsection{Misalignment mechanism}
First, we focus on the production of the ALP through the misalignment mechanism~\cite{Preskill:1982cy, Abbott:1982af, Dine:1982ah}, where the initial field displacement from its potential minimum in the early universe leads to coherent oscillations that contribute to the present-day dark matter abundance. We derive the relic abundance and discuss the parameter space consistent with current observational constraints. The subsequent evolution of the homogeneous ALP field is governed by the equation of motion in an expanding universe as
\begin{align}
\ddot \phi + 3 H \dot \phi + m_\phi^2 \phi = 0,    
\end{align}
where $H$ is the Hubble parameter and $m_\phi$ is the ALP mass. At early times when $H > m_\phi$, the Hubble friction effectively freezes the field, while at later times when $H \lesssim m_\phi$, the field begins coherent oscillations around the potential minimum. These oscillations behave as non-relativistic matter.

Let us define the temperature of the onset of the oscillation, $T_{\rm osc}$, as $3H(T_{\rm osc}) = m_\phi$. Then, by using $H(T) = 1.66 \sqrt{g_*(T)} (T^2/M_{\rm Pl})$, we obtain
\begin{align}
    T_{\rm osc} \sim \sqrt{m_\phi M_{\rm Pl}} \sim 10^6~{\rm GeV} \times \left( \frac{m_\phi}{10~{\rm keV}} \right)^{1/2}.
\end{align}
Note that $g_{*}(T)$ denotes the effective number of relativistic degrees of freedom for energy density and $M_{\rm Pl} = 1.22 \times 10^{19} \, \mathrm{GeV}$.

Depending on the relative magnitude of the reheating temperature $T_{\rm RH}$ and the temperature at which the ALP field begins coherent oscillations $T_{\rm osc}$, the misalignment mechanism can operate in two distinct cosmological regimes.
The first possibility corresponds to $T_{\rm RH} < T_{\rm osc}$, in which the onset of ALP oscillations occurs before the completion of reheating. In this case, the Universe is still dominated by the oscillating inflaton condensate, and the ALP starts oscillating during an effectively matter-dominated epoch. In this regime, the expansion history is modified, and the relation $H(T)=1.66\sqrt{g_*}\,T^2/M_{\rm Pl}$ is no longer valid at the onset of oscillations. Also, the continuous decay of the inflaton produces entropy, which dilutes the comoving ALP number density after oscillations have begun. Consequently, the final ALP abundance depends not only on the ALP parameters $(m_\phi,f,\theta_0)$, but also on the details of the reheating process, including the reheating temperature and the inflaton decay dynamics. Eventually, for a decay constant relevant for generic freeze-in scenarios and the lowest reheating temperatures allowed by BBN, the ALP relic abundance in this regime is given by~\cite{Blinov:2019rhb},
\begin{align}
	\Omega_\phi h^2
    \sim 1.48 \times 10^{-11} \left( \frac{f \,\theta_0}
	{10^{10}~{\rm GeV}} \right)^2 \times \left(\frac{T_{\rm RH}}{10~{\rm MeV}}\right),
\end{align}
In the present work, we focus on the second regime, $T_{\rm RH} > T_{\rm osc}$, where reheating is completed before the onset of ALP oscillations. In this case, the Universe is already radiation-dominated when the ALP field starts oscillating, entropy is conserved thereafter, and the relic abundance can be computed in a model-independent manner without specifying the details of the reheating history. The resulting abundance depends only on the ALP parameters and standard cosmological quantities, making this scenario particularly suitable for our analysis.

In the case of $T_{\rm RH} > T_{\rm osc}$, the relic abundance of the ALP can be expressed in terms of $m_{\phi}$ and $f$. The current energy density of ALP $\rho_{\phi,0}$ is given by~\cite{Arias:2012az}
\begin{equation}
\rho_{\phi, 0} = \frac{1}{2} m_\phi^2  \phi_{\rm osc}^2 \frac{g_{*s} (T_0)T_0^3}{g_{*s} (T_{\rm{osc}})  T_{\rm{osc}}^3} ,
\label{eq:rho_phi}
\end{equation}
where $\phi_{\rm osc} = f \theta_0$ is the field amplitude at the onset of oscillations and $\theta_0$ is the corresponding angle, 
$g_{*s}(T)$ represents the effective number of entropy degrees of freedom, and $T_0 \simeq 2.7~{\rm K}$.
Finally, we obtain
\begin{align}
    \Omega_\phi h^2|_{\rm mis} &= \frac{\rho_{\phi,0}}{\rho_{c,0}}h^2 = 8\times 10^{-3} \sqrt{\frac{m_\phi}{10~\rm{keV}}}   \qty(\frac{f\theta_0}{10^{10}~\rm{GeV}})^2 \mathcal{F}(T_{\rm{osc}}), \label{eq:omegah2_MM}
\end{align}
where $\mathcal{F}(T_{\rm{osc}}) \equiv (g_*(T_{\rm{osc}})/106.75)^\frac{3}{4} (g_{*s}(T_{\rm{osc}})/106.75)^{-1} $
and $\rho_{c,0}\simeq 1.053\times 10^{-5}~h^2 ~\rm{GeV}~\rm{cm^{-3}}$. 

A comment on the range of the initial misalignment angle is in order. Because the accidental $U(1)$ is broken to $Z_N$, the potential has period $2\pi f/N$ and the displacement from the nearest minimum obeys $|\theta_0| \le \pi/N$, with $\theta_0 \equiv \phi_{\rm osc}/f$. Rather than fixing $\theta_0$ by hand, we take for each $N$ the untuned value $\theta_0 = \pi/(N\sqrt{3})$, the root-mean-square of a uniformly distributed angle over the allowed range, which is the relevant average since $\Omega_\phi h^2 \propto \theta_0^2$. Figure~\ref{fig:benchmarks} shows the resulting contours for $N = 5,\,8,\,11$.

\subsubsection{Freeze-in mechanism}
\label{sec:direct_freezein}

The derivative interactions of eqs.~\eqref{eq:ALP quark coupling type-D} and \eqref{eq:ALP quark coupling type-U} provide a natural mechanism for the thermal production of the ALP through freeze-in~\cite{Hall:2009bx}. For sufficiently large values of the symmetry-breaking scale $f$, the ALP interaction rate remains below the Hubble rate throughout the thermal history of the Universe: the ALP never thermalizes with the Standard Model plasma, and a relic population is slowly accumulated through rare decays and scatterings of bath particles. In this section we spell out the cosmological assumptions under which this computation is well defined, derive the production rates in the two relevant thermal regimes, describe how they are matched across the QCD transition, and quantify the theoretical uncertainties of the procedure. As we will see, the individual production rates carry order-one ambiguities inherited from strong dynamics near the GeV scale; we assess each of them explicitly and show that, because the relic abundance determines the freeze-in contour only through $f_{\rm FI}\propto\sqrt{Y_\phi}$ and the parameter space of interest spans many decades in $f$, these uncertainties do not affect the conclusions of this work, significantly.

Freeze-in calculations are notoriously sensitive to the highest temperature of the thermal bath whenever the production rates grow with temperature. In the Nelson--Barr context, this question acquires a structural dimension, because the theory possesses a physical UV scale, $f$, at which CP is spontaneously broken, and the heavy vector-like quarks live. If the reheating temperature exceeds this scale, $T_{\rm RH}\gtrsim f$, the heavy sector is thermally populated and the ALP abundance receives UV-dominated contributions from its decays and scatterings; the result then depends on $T_{\rm RH}$ and on the detailed spectrum of the CP-breaking sector, and the predictivity of the low-energy analysis is lost. More importantly, this regime is theoretically disfavored on grounds independent of dark matter. At $T\gtrsim f$ thermal corrections generically restore the spontaneously broken CP symmetry; its subsequent re-breaking produces cosmological domain walls, CP being a discrete symmetry, which must be inflated away or otherwise removed. In addition, a thermally populated CP-breaking sector can displace the fields responsible for the Nelson--Barr texture, reintroducing contributions to $\bar\theta$ and threatening the very solution of the strong CP problem that motivates the construction.

We therefore assume $T_{\rm RH}<f$ throughout. This choice ensures that the CP-breaking vacuum selected at the end of inflation is never thermally erased, keeps the heavy Nelson--Barr states out of equilibrium, and renders the ALP production computable entirely within the low-energy EFT of eqs.~\eqref{eq:ALP quark coupling type-D} and \eqref{eq:ALP quark coupling type-U}.
First, let us discuss the UV freeze-in contributions in such a situation. By using the EOM of quarks, the effective interaction in the type-D model given in eq.~\eqref{eq:ALP quark coupling type-D} can be written as
\begin{align}
    {\cal L } &\ni -i \frac{y_{d,ik} C^{(D)}_{kj} }{f} \phi H q_{L,i} \widehat{d}_{Rj} + h.c..
\end{align}
We can derive a similar interaction for the type-U model. This dimension-5 operator gives a larger production rate at high temperature \cite{Hall:2009bx}. By using formulae in ref.~\cite{Aghaie:2024jkj}, we estimate its contribution\footnote{The details of the relic abundance computation are demonstrated explicitly in Appendix F of Ref.~\cite{Ziegler:2026kis}.} as,
\begin{align}
    \Omega_a h^2 |_{\rm UV} \sim 0.1
    \times \left(\frac{T_{\rm RH}}{3 \times 10^6~{\rm GeV}} \right)
    \left( \frac{m_a}{0.1~{\rm MeV}} \right)
    \left( \frac{10^{10}~{\rm GeV}}{f_a}\right)^2
    \left( \frac{m_{q_i}}{1~{\rm GeV}}\right)^2
    \left| C_{ij} \right|^2.
\end{align}
Note that $q_i$ in the above formula should be a heavy quark, i.e., the bottom quark for the type-D model, and the top quark for the type-U model.
As we will see in section \ref{sec:constraints}, the coupling to the heavy quark is suppressed in most of the parameter space because of the structure of the CKM matrix. Thus, the UV freeze-in contribution is naturally suppressed. In this paper, we focus on a scenario such that the UV freeze-in mechanism is not enough to give the relic abundance of the dark matter.
In this case, the IR freeze-in mechanism should give the dominant contributions.
Crucially, no fine-grained knowledge of $T_{\rm RH}$ is needed: all production channels considered below are IR-dominated, i.e., decay rates peak at $T\sim m_q/3$, and the scattering cross sections carry explicit $m_q^2/s$ chirality suppressions, so the relic abundance is insensitive to the reheating temperature for any $m_q \ll T_{\rm RH} < f$. The freeze-in prediction thus depends only on measured quark masses, the coupling matrices $C_{ij}$ fixed by the Nelson--Barr flavor structure, and the scale $f$.

\paragraph{Two thermal regimes and the QCD crossover.}
Below $f$ the relevant hierarchy of scales is
\begin{equation}
	m_\phi \ll \Lambda_{\rm QCD} \ll f.
\end{equation}
As the Universe cools through the strong-interaction scale, the degrees of freedom carrying the ALP couplings reorganize. Lattice QCD establishes that at zero baryon chemical potential this reorganization is not a phase transition but an analytic \emph{crossover}, with a pseudo-critical temperature $T_{\rm pc}\simeq\Tpc$~\cite{Aoki:2006we}: there is no sharp temperature at which quarks and gluons cease to be the appropriate description and hadrons take over. Any freeze-in computation spanning this epoch must therefore adopt a \emph{matching prescription} between a partonic description, valid at $T\gg T_{\rm pc}$, and a hadronic description, valid at $T\ll T_{\rm pc}$---and the prescription dependence must be quantified as a theoretical uncertainty. Our strategy, detailed below, is to partition the thermal history at a switching temperature $T_{\rm sw}$ and to propagate the ambiguities of this choice, together with the intrinsic limitations of the two effective descriptions, into an uncertainty band on the freeze-in contour. We emphasize at the outset that the present work is not a precision relic-abundance calculation: our aim is to establish where in the $(m_\phi,f)$ plane freeze-in production is viable, and, as quantified below, the residual uncertainties shift the corresponding contours by tens of percent in $f$, which is only a small fraction of a decade in the parameter space.

\paragraph{Boltzmann framework.}
Since the ALP population remains far below equilibrium, inverse processes can be neglected and the Boltzmann equation for the yield $Y_\phi=n_\phi/s$ linearizes,
\begin{equation}
	\frac{dY_\phi}{dT} = -\frac{1}{s\,H\,T}
    \left( 1 + \frac{1}{3} \frac{d\log g_{*s}}{d\log T}\right)
	\Big[ \textstyle\sum_{i,j}\, \gamma_{\rm dec}(q_i\to q_j\phi)
	+ \sum_{ab}\, \gamma(ab\to c\,\phi) \Big],
	\label{eq:boltzmann_FI}
\end{equation}
with decay and scattering reaction densities
\begin{equation}
	\gamma_{\rm dec} = \frac{g_i\, m_i^2\, \Gamma\, T}{2\pi^2}\,
	K_1\!\left(\frac{m_i}{T}\right),
	\qquad
	\gamma(ab\to c\,\phi) = \frac{T}{64\pi^4}
	\int_{s_{\rm min}}^{s_{\rm max}} \! ds\, \sqrt{s}\;
	\widehat\sigma(s)\, K_1\!\left(\frac{\sqrt{s}}{T}\right),
	\label{eq:gamma_def}
\end{equation}
where $g_i$ counts the internal degrees of freedom of the decaying species, and $\widehat\sigma(s)$ is the reduced cross section defined as,
\begin{equation}\widehat\sigma(s)=\frac{2\lambda(s,m_a^2,m_b^2)\,\sigma(s)}{s}, \quad {\rm with} \quad \lambda(s,m_a^2,m_b^2)=\qty(s-(m_a+m_b)^2)\qty(s-(m_a-m_b)^2),
\end{equation}
where $\lambda(s,m_a^2,m_b^2)$ is the K\"all\'en function. We integrate eq.~\eqref{eq:boltzmann_FI} numerically for each channel, using the lattice equation of state for the effective degrees of freedom $g_*(T)$, $g_{*s}(T)$ across the crossover~\cite{Saikawa:2020swg}. For decays, the familiar closed form~\cite{Hall:2009bx},
\begin{equation}
	Y_\phi^{\rm dec} \simeq \frac{135}{8\pi^3(1.66)}
	\frac{g_q}{g_{*s}\sqrt{g_*}}
	\frac{M_{\rm Pl}\,\Gamma(q_i\to q_j\phi)}{m_{q_i}^2},
	\label{eq:Ydec_freezein}
\end{equation}
reproduces our numerical decay yields to good accuracy. Because every rate scales as $1/f^2$, the abundance of any channel or category can be summarized by a single number: the decay constant $f_\star$ at which that channel alone would saturate the observed abundance, $\Omega_\phi h^2 = 0.12\,(m_\phi/\mbench)(f_\star/f)^2$. Large $f_\star$ identifies an efficient production mode.

\paragraph{Partonic production ($T>T_{\rm sw}$).}
Above the crossover, the ALP couples to quarks and gluons. The dominant channels are the flavor-changing decays $q_i\to q_j\phi$ with width
\begin{equation}
	\Gamma(q_i\to q_j\phi) = \frac{m_{q_i}^3}{32\pi f^2}\,|C_{ij}|^2
	\left(1-\frac{m_{q_j}^2}{m_{q_i}^2}\right)^{\!3},
	\label{eq:Gamma_qtoqphi}
\end{equation}
where $C_{ij}$ stands for $C^{(D)}_{ij}$ or $C^{(U)}_{ij}$: the leading decays are $b\to d\phi,\,s\phi$ in the type-D model and $t\to u\phi$, $c\to u\phi$ in the type-U model. The diagonal couplings additionally induce the QCD-assisted scatterings $q_i g\to q_i\phi$ and $q_i\bar q_i\to g\phi$, with tree-level cross sections~\cite{Aghaie:2024jkj}
\begin{align}
	\sigma_{q_i g\to q_i\phi} &= \frac{\alpha_s}{48 f^2}\,|C_{ii}|^2\,
	\frac{x}{1-x}\left(-2\ln x - 3 + 4x - x^2\right), \\
	\sigma_{q_i\bar q_i\to g\phi} &= \frac{4\alpha_s}{9 f^2}\,|C_{ii}|^2\,
	x\,\frac{\tanh^{-1}\!\sqrt{1-4x}}{1-4x},
\end{align}
with $x=m_{q_i}^2/s$. The analogous photon-induced channels, obtained by $\alpha_s/6\to\alpha_{\rm em}Q_i^2$ and $4\alpha_s/9\to\alpha_{\rm em}Q_i^2$, are suppressed by $\alpha_{\rm em}Q_i^2/\alpha_s\ll 1$ and neglected. Two comments on the domain of validity are in order. First, for the light flavors the kinematic thresholds $\sqrt{s}\sim m_{d},m_s$ lie deep in the nonperturbative regime, where the partonic cross sections and the running of $\alpha_s$ are meaningless; we therefore evaluate all partonic scattering rates only for $\sqrt{s}>\pertFloor$. Sub-GeV collisions are instead described by the hadronic channels below, and the residual dependence on this prescription is part of the matching uncertainty. Second, the decay channels are exempt from these subtleties: $b$-quark and heavy up-sector decays complete at $T\sim m_q/3\gg T_{\rm pc}$, far above the crossover, a fact that will be reflected in the stability of their yields below.

The off-diagonal couplings also generate flavor-violating scatterings, $b\bar q\to\phi g$, $bg\to\phi q$ and $qg\to\phi b$ ($q=d,s$), controlled by the same $|C_{bd}|,|C_{bs}|$ that fix the decays $b\to q\phi$. These processes are individually infrared-divergent in the $m_q\to0$ limit, however, as it is shown in ref.~\cite{Aghaie:2024jkj}, the collinear singularities cancel by the Kinoshita--Lee--Nauenberg theorem through the anomalous-threshold cuts, while the soft-gluon divergence is removed by the thermal-mass and wave-function corrections to the $b\to q\phi$ decay, leaving a finite scattering-to-decay ratio $\gamma_S/\gamma_D\simeq0.7$ at temperatures $T\sim m_b/3$ where production peaks. We do not recompute these contributions here; including them would rescale the (subdominant) flavor-violating quark-decay category by an $\mathcal{O}(1)$ factor, shifting the total type-D yield by $\lesssim20\%$ and the freeze-in contour by $\lesssim10\%$ in $f$---within the matching uncertainty of table~\ref{tab:Tsw_scan} and immaterial for the type-U model, where the flavor-violating couplings are hierarchically suppressed. We stress that, in both models, the dominant production is pion scattering below the crossover, which is unaffected.

\paragraph{Hadronic production ($T<T_{\rm sw}$).}
Below the crossover, the ALP couplings are carried by mesons, as encoded in the chiral effective theory of Appendix~\ref{app:meson_eft}. The available channels follow from the flavor structure. In the type-U model the matching gives $C'^{(U)}=\mathrm{diag}(C_{uu},0,0)$: all flavor-changing ALP--meson vertices vanish, and no mesonic decays arise. In the type-D model, the coupling $C_{sd}$ generates the rare decays $K^0,\bar K^0\to\pi^0\phi$. In both models, the cubic term of the chiral current contains a four-point vertex of the ALP with three pions,
\begin{equation}
	\mathcal L_{\phi\pi\pi\pi} = -\frac{C_{dd}-C_{uu}}{3 f f_\pi}\,
	\partial_\mu\phi\left[\pi^0\pi^+\partial^\mu\pi^-
	+ \pi^0\pi^-\partial^\mu\pi^+ - 2\pi^+\pi^-\partial^\mu\pi^0\right],
	\label{eq:phi3pifreezein}
\end{equation}
nonvanishing in both realizations ($C_{uu}=0$ in type-D, $C_{dd}=0$ in type-U), which induces $\pi^+\pi^-\to\pi^0\phi$ and $\pi^\pm\pi^0\to\pi^\pm\phi$. The amplitude of the neutral channel is $\mathcal M=-\tfrac{C_{dd}-C_{uu}}{2ff_\pi}(s-m_\pi^2)$, giving
\begin{equation}
	\sigma(\pi^+\pi^-\to\pi^0\phi) =
	\frac{(C_{dd}-C_{uu})^2}{64\pi f^2 f_\pi^2}\,
	\frac{(s-m_\pi^2)^3}{s\,\lambda^{1/2}(s,m_\pi^2,m_\pi^2)}.
	\label{eq:pipitopiphi}
\end{equation}
Unlike the two-meson vertices, whose on-shell matrix elements vanish for degenerate mesons, this vertex produces a nonvanishing rate at leading order, and pions are the most abundant hadrons throughout the confined epoch, so the associated reaction density is only mildly Boltzmann suppressed. At the same order the chiral current generates $\phi\pi K\bar K$ and $\phi\eta^{(\prime)}K\bar K$ vertices, inducing $K\bar K\to\pi^0\phi,\,\eta\phi,\,\eta'\phi$---through $K^+K^-$ in the type-U model and dominantly through $K^0\bar K^0$ in the type-D model---suppressed by two kaon Boltzmann factors.

The leading-order chiral amplitudes grow with energy, i.e., the cross section \eqref{eq:pipitopiphi} scales as $s$ at large $s$, which is nothing but the breakdown of the momentum expansion: beyond $\Lambda_\chi = 4\pi f_\pi\simeq\LchiVal$ the amplitudes violate unitarity and carry no physical information. Left uncut, this spurious growth would overwhelm the Boltzmann suppression of the high-energy tail and dominate the reaction density which is a purely unphysical effect. We therefore evaluate all hadronic reaction densities only for $\sqrt{s}<\Lambda_\chi$, and treat the precise location of the cutoff as a source of uncertainty. We do not attempt to model the resonance region ($\rho$, $K^*$, \dots) that unitarizes the true amplitudes above $\Lambda_\chi$; its net effect is bracketed by the cutoff variation below.

\paragraph{Matching and theoretical uncertainties.}
The two descriptions are combined by partitioning the temperature integral of eq.~\eqref{eq:boltzmann_FI} at a switching temperature $T_{\rm sw}$: partonic rates for $T>T_{\rm sw}$, hadronic rates for $T<T_{\rm sw}$. Each temperature interval contributes exactly once, so the prescription is free of double counting by construction. The price is a set of systematic ambiguities, which we enumerate together with our handling of each:
\begin{itemize}
\item[(i)] \emph{Choice of $T_{\rm sw}$.} Since the crossover is analytic, $T_{\rm sw}$ is not a physical temperature; we take $T_{\rm sw}=T_{\rm pc}\simeq\Tpc$ as central value and vary it over $\TswWin$.
\item[(ii)] \emph{Chiral cutoff.} The hadronic rates depend on where the $\sqrt{s}$ integration is terminated; we vary $\Lambda_\chi\in\LchiWin$.
\item[(iii)] \emph{Perturbative floor.} The light-flavor partonic rates depend mildly on the $\sqrt{s}>\pertFloor$ prescription; this variation is subleading to (i) and (ii) because those channels contribute at or below the permille level (cf.\ Table~\ref{tab:contributions}).
\item[(iv)] \emph{Equation of state.} $g_*(T)$ varies rapidly across the crossover; we use the lattice tabulation of ref.~\cite{Saikawa:2020swg} throughout, which renders this a negligible source of error compared to (i) and (ii).
\end{itemize}
Table~\ref{tab:Tsw_scan} displays the $T_{\rm sw}$ dependence of each production category. Two features bear emphasis. First, the quark-decay yields are independent of $T_{\rm sw}$ at the permille level---as they must be, since heavy-quark decays complete far above the crossover; this constitutes a nontrivial internal consistency check of the matching. Second, the mesonic contributions are generated in the immediate vicinity of the matching point, and roughly double between $T_{\rm sw}=120$ and $200~{\rm MeV}$: they are the dominant carriers of the matching uncertainty. The total yield consequently varies by a factor of order two across the window---but the freeze-in contour responds only through $f_{\rm FI}\propto\sqrt{Y_\phi}$, shifting by $\DbandTsw$ (type-D) and $\UbandTsw$ (type-U). On the logarithmic $(m_\phi,f)$ plane of figure~\ref{fig:benchmarks} these shifts amount to less than a tenth of a decade, to be compared with viability windows and exclusion depths that extend over several decades; none of the conclusions drawn in our analysis is sensitive to variations of this size. The same comment applies to the composition uncertainties discussed next.

\begin{table}[t]
	\renewcommand{\arraystretch}{1.35}
	\setlength{\tabcolsep}{4pt}
	\centering
	\small
	\begin{tabular}{l c c c c c c c}
		\hline\hline
		& $T_{\rm sw}$ [GeV] & Quark dec. & Quark scatt. & Meson dec. &
		Meson scatt. & Total & $f_{\rm FI}/f_{\rm FI}^{155}$ \\
		\hline
		\multirow{3}{*}{Type-U}
		& $0.120$ & $6.42\times10^{10}$ & $5.7\times10^{8}$ & --- & $2.36\times10^{13}$ & $2.37\times10^{13}$ & $0.70$ \\
		& $0.155$ & $6.26\times10^{10}$ & $5.7\times10^{8}$ & --- & $4.76\times10^{13}$ & $4.77\times10^{13}$ & $1$ \\
		& $0.200$ & $5.93\times10^{10}$ & $5.7\times10^{8}$ & --- & $7.54\times10^{13}$ & $7.55\times10^{13}$ & $1.26$ \\
		\hline
		\multirow{3}{*}{Type-D}
		& $0.120$ & $1.26\times10^{13}$ & $3.6\times10^{9}$ & $3.03\times10^{12}$ & $2.25\times10^{13}$ & $3.82\times10^{13}$ & $0.78$ \\
		& $0.155$ & $1.26\times10^{13}$ & $3.5\times10^{9}$ & $4.47\times10^{12}$ & $4.55\times10^{13}$ & $6.25\times10^{13}$ & $1$ \\
		& $0.200$ & $1.26\times10^{13}$ & $3.3\times10^{9}$ & $5.43\times10^{12}$ & $7.21\times10^{13}$ & $9.01\times10^{13}$ & $1.20$ \\
		\hline\hline
	\end{tabular}
	\caption{Dependence of the freeze-in yields on the switching
	temperature $T_{\rm sw}$ in the two models, at the respective benchmarks
	with $f=1$~GeV (all rates scale as $1/f^2$). The quark-decay contributions
	are $T_{\rm sw}$-independent at the permille level, while the mesonic
	contributions, produced near the matching point, vary by a factor of three
	across the window. Mesonic decays are absent in the type-U model (see text).
	The last column gives the induced shift of the freeze-in contour,
	$f_{\rm FI}\propto\sqrt{Y_\phi}$, relative to $T_{\rm sw}=155$~MeV.}
	\label{tab:Tsw_scan}
\end{table}

\paragraph{Results at the benchmarks.}
Table~\ref{tab:contributions} decomposes the relic abundance at the two benchmark points, obtained by solving eq.~\eqref{eq:boltzmann_FI} with all channels included. The observed abundance is reproduced for $f=\DfFItotal$ (type-D) and $f=\UfFItotal$ (type-U) at $m_\phi=\mbench$.

The most striking feature of both tables is that \emph{hadronic production dominates over partonic production in both models}: mesonic channels account for \DshareHadronic\ of the type-D and \UshareHadronic\ of the type-U abundance, with the single process $\pi^+\pi^-\to\pi^0\phi$ contributing \DsharePiPi\ and \UsharePiPi\ respectively. This outcome is far from obvious. Naively, the heavy-quark decay channels, with widths scaling as $m_q^3/f^2$ and production active over many Hubble times, would be expected to dominate over processes involving only sub-GeV states active during a brief window of the thermal history. That they do not is a direct imprint of the Nelson--Barr flavor structure: the flavor-changing couplings that drive quark decays are built from products of heavy--light mixings and are strongly suppressed, while the diagonal light-quark couplings that feed the chiral vertex \eqref{eq:phi3pifreezein} are unsuppressed. The models then differ in the composition of the partonic remainder. In type-D, the decays $b\to d\phi,\,s\phi$ retain a sizable share (\DshareQuarkDecay\ in total, with $b\to d\phi$ alone at \DshareBtoD), the $C_{sd}$-induced kaon decays add \DshareMesonDecay, and quark scattering is suppressed to \DshareQuarkScatt. In type-U, the flavor suppression is more severe: the partonic total is only \UshareQuarkDecay\ (dominated by $t\to u\phi$), mesonic decays are forbidden outright, and freeze-in is essentially saturated by pion scattering.

\begin{table}[t]
\centering
\renewcommand{\arraystretch}{1.4}
\setlength{\tabcolsep}{6pt}
\begin{tabular}{c c c c c c}
\hline\hline
& \textbf{Process category} & \textbf{Main channel} & \textbf{Yield} &
\textbf{Percent} & \textbf{$f_\star$ [GeV]} \\
\hline
\multirow{4}{*}{Type-U}
& Meson scattering & $\pi^+\pi^- \to \pi^0\phi$      & $4.34\times10^{-7}$ & $99.87\%$             & $1.05\times10^{10}$ \\
& Quark decay      & $t \to u\phi$                   & $5.71\times10^{-10}$ & $0.13\%$             & $3.8\times10^{8}$ \\
& Quark scattering & $u\,g \to u\,\phi$              & $5.24\times10^{-12}$ & $1.2\times10^{-3}\,\%$ & $3.6\times10^{7}$ \\
& Meson decay      & ---                             & $0$                 & $0$                   & --- \\
\hline
\multirow{4}{*}{Type-D}
& Meson scattering & $\pi^+\pi^- \to \pi^0\phi$      & $3.16\times10^{-7}$ & $72.7\%$              & $1.02\times10^{10}$ \\
& Quark decay      & $b \to d\phi$                   & $8.76\times10^{-8}$ & $20.1\%$              & $5.4\times10^{9}$ \\
& Meson decay      & $K^\pm \to \pi^\pm\phi$         & $3.11\times10^{-8}$ & $7.1\%$               & $3.2\times10^{9}$ \\
& Quark scattering & $d\,g \to d\,\phi$              & $2.42\times10^{-11}$ & $0.006\%$            & $8.9\times10^{7}$ \\
\hline\hline
\end{tabular}

\caption{Contributions to the dark-matter relic density by process category in the type-D and type-U models, at the respective benchmarks ($m_\phi=1~{\rm MeV}$, $T_{\rm sw}=155~{\rm MeV}$, $\sqrt{s}<\Lambda_\chi$ in the hadronic channels). For each category, the dominant channel is shown; the last column gives the decay constant $f_\star$ at which that category alone would reproduce the observed abundance. Mesonic decays are absent in the type-U model, where the low-energy coupling matrix is purely diagonal.}
\label{tab:contributions}
\end{table}

Two qualifications frame this observation. First, it is a statement about the benchmark points: the shares depend on the coupling matrices $C_{ij}$, which vary across the Nelson--Barr parameter space scanned in section~\ref{sec:robustness}, and configurations in which quark decays compete with or exceed the pion channel do occur. What is general is the lesson that the hadronic channels can dominate and must be included, i.e. a purely partonic computation would have missed the leading production mode at both benchmarks and misplaced the freeze-in contour accordingly. Second, the prominence of $\pi\pi\to\pi\phi$ has a well-known analogue in thermal QCD axion production, where $\pi\pi\leftrightarrow\pi a$ scattering controls the axion's thermalization and abundance near the crossover~\cite{Bauer:2020jbp,DiLuzio:2022gsc}, and where the treatment of precisely the region $T\sim T_{\rm pc}$ has been the subject of considerable scrutiny. The parallel is instructive but not exact: there the vertex descends from the anomalous gluon coupling, whereas here it originates from the anomaly-free derivative couplings dictated by the Nelson--Barr flavor structure, the ALP of this work couples to the chiral currents, not to $G\widetilde G$. The shared feature is structural: for any light pseudoscalar coupled to first-generation quarks, the pion bath is the last, and often the most efficient, thermal reservoir available.

\paragraph{Validity of the freeze-in computation.}
The linearization of eq.~\eqref{eq:boltzmann_FI} assumes the ALP population remains far below equilibrium. Since $Y_\phi\propto 1/f^2$, this fails below a critical decay constant at which $Y_\phi$ reaches $Y_\phi^{\rm eq}=45\zeta(3)/(2\pi^4 g_{*s})\simeq 0.28/g_{*s}$ and the ALP thermalizes with the bath.\footnote{This integrated criterion is stronger than the naive comparison of the vacuum width with $H(m_q)$ by roughly an order of magnitude, because production accumulates over several Hubble times.} Solving $Y_\phi(f_{\rm th})=Y_\phi^{\rm eq}$ gives $f_{\rm th}\simeq\fthD$ (type-D) and $f_{\rm th}\simeq\fthU$ (type-U): for $f\lesssim f_{\rm th}$ the ALP thermalizes independently of $m_\phi$, and the freeze-in computation is void, the relic would instead be a thermal population, excluded as hot dark matter for the masses of interest. All freeze-in results in this work therefore apply only for $f\gtrsim f_{\rm th}$; we dash the corresponding portion of the contours in figures~\ref{fig:benchmarks} and \ref{fig:margins}.

\begin{figure}[p]
  \centering
  \includegraphics[width=0.9\textwidth]{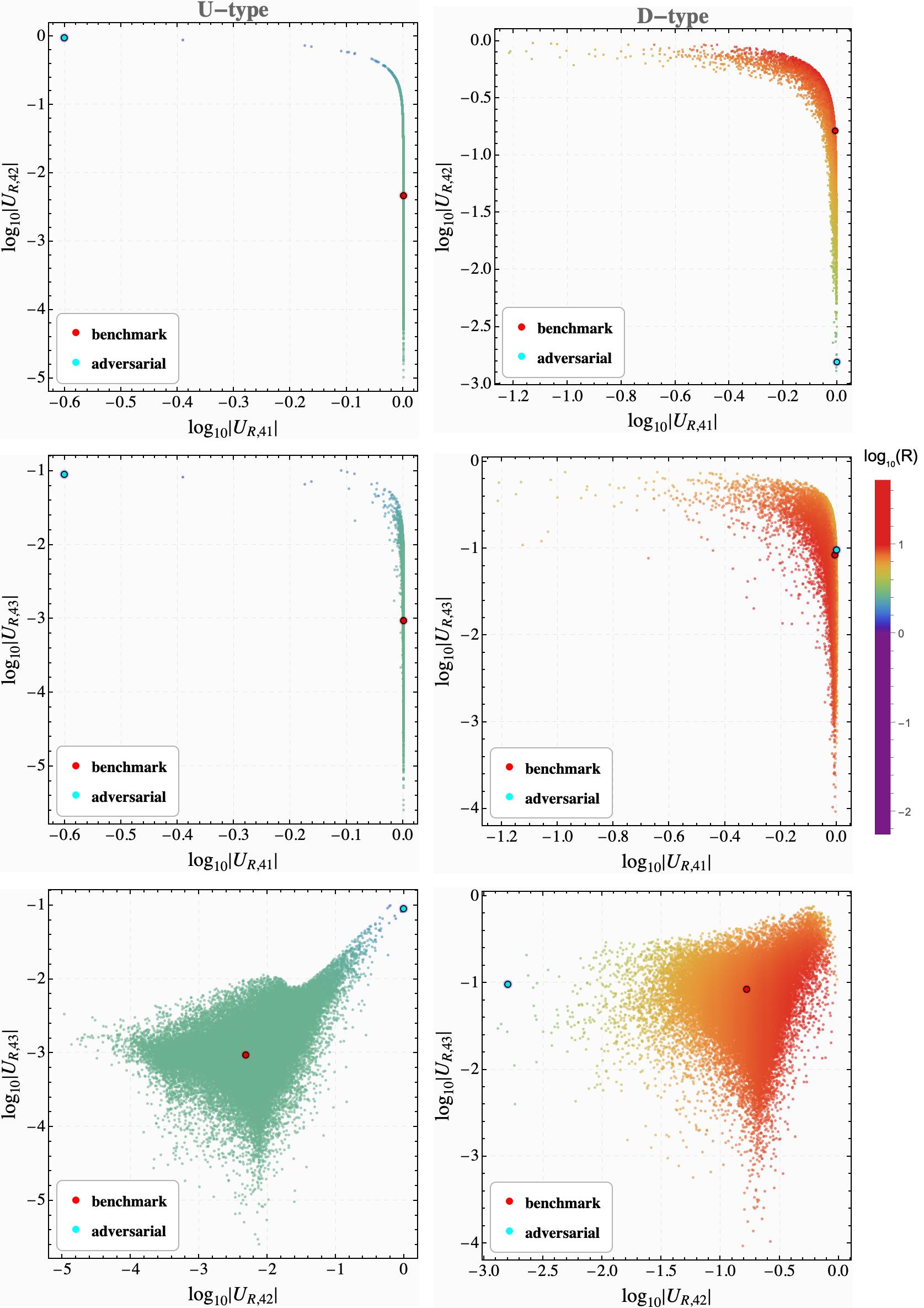}
  \caption{Scan of $\Nscan$ points over the free parameters $(b,\gamma,\beta_2,\beta_3)$, shown in the planes of the right-handed mixing elements $|U_{R,4i}|$ for the type-U (left column) and type-D (right column) models. Each point is colored by its exclusion margin $\mathcal{R}$, defined in eq.~\eqref{eq:margin} as the minimum over $m_\phi$ of the ratio $f_{\rm excl}/f_{\rm FI}$ between the strongest current bound and the freeze-in contour: points with $\mathcal{R}>1$ have their entire freeze-in contour excluded, while for $\mathcal{R}<1$ a mass window remains in which freeze-in production evades all current constraints. The red marker denotes the representative benchmark used throughout the text, introduced in eq.~\eqref{eq:benchmark_couplings}, the cyan marker denotes the \emph{adversarial} benchmark, selected as the scan point minimizing $\mathcal{R}$ and therefore the most favorable to freeze-in.}
  \label{fig:scan}
\end{figure}

\subsection{Constraints and future prospects}
\label{sec:constraints}

As discussed in section~\ref{sec:model}, once the quark masses and the CKM matrix are fixed, the coupling matrices $C^{(D)}$ and $C^{(U)}$ are fully determined by the four Nelson--Barr parameters $(b,\gamma,\beta_2,\beta_3)$. Figure~\ref{fig:scan} shows the scatter plots of $U_{R,4i}$ for the type-D and type-U model. For our analysis, we take the SM quark masses as $m_u = 2.16~{\rm MeV}$, $m_d = 4.70~{\rm MeV}$, $m_s = 93.5~{\rm MeV}$, $m_c = 1.27~{\rm GeV}$,  $m_b = 4.18~{\rm GeV}$, and $m_t = 173~{\rm GeV}$ \cite{ParticleDataGroup:2024cfk}. For the CKM parameters, we take $\sin\theta_{12} = 0.225$, $\sin\theta_{13} = 0.00373$, $\sin\theta_{23} = 0.0418$, and $\delta = 1.15$ \cite{ParticleDataGroup:2024cfk}.  The scatter plot in figure~\ref{fig:scan} reveals the hierarchy $|U_{R,43}| \lesssim |U_{R,42}| \lesssim |U_{R,41}|$ in most of the parameter space in both of the type-D and type-U models. For simplicity, we neglect the RG running effect on the SM Yukawa couplings between the electroweak scale and the Nelson--Barr scale. Although the RG effect could change ${\cal O}(1)$ factor in the discussion, it does not change the order of the lifetime of the ALP and the flavor structure discussed later.

To present the constraints, we adopt a common benchmark point for both models,
\begin{equation}
        b = 0.20, \qquad \gamma = 1.0, \qquad \beta_2 = \frac{\pi}{2}, \qquad \beta_3 = \frac{\pi}{4}.
        \label{eq:benchmark_params}
\end{equation}
This benchmark is chosen for illustrative purposes only and there is nothing special about it. As we show in section~\ref{sec:robustness}, every constraint and every freeze-in rate depends on the parameters only through the modulus of a single coupling combination, cf.~eq.~\eqref{eq:factorization}, and these moduli vary by ${\cal O}(1)$ factors over the parameter space, so no qualitative conclusion rests on this benchmark; section~\ref{sec:robustness} promotes all statements of this subsection to a systematic scan over $(b,\gamma,\beta_2,\beta_3)$.

Since none of the observables considered below is sensitive to the phases of the off-diagonal couplings, i.e., flavor-violating rates scale as $|C_{ij}|^2$ and the induced photon coupling involves only the real diagonal entries, it suffices to quote the moduli of the coupling matrices at the benchmark point. In the flavor bases $(d,s,b)$ and $(u,c,t)$ respectively,
\begin{equation}
        \big|C^{(D)}\big| =
        \begin{pmatrix}
                0.964 & 0.163 & 0.0843 \\
                0.163 & 0.0276 & 0.0143 \\
                0.0843 & 0.0143 & 0.00737
        \end{pmatrix},
        \quad
        \big|C^{(U)}\big| =
        \begin{pmatrix}
                1.00 & 4.71\times10^{-3} & 9.51\times10^{-4} \\
                4.71\times10^{-3} & 2.22\times10^{-5} & 4.48\times10^{-6} \\
                9.51\times10^{-4} & 4.48\times10^{-6} & 9.04\times10^{-7}
        \end{pmatrix}.
        \label{eq:benchmark_couplings}
\end{equation} The structure of both matrices is inherited from the Nelson--Barr Yukawa construction rather than from the parameter choice: the diagonal entries obey $C_{dd}^{(D)} : C_{ss}^{(D)} : C_{bb}^{(D)} \approx 1 : 0.029 : 0.0076$ and $C_{uu}^{(U)} : C_{cc}^{(U)} : C_{tt}^{(U)} \approx 1 : 2.2\times10^{-5} : 9.0\times10^{-7}$, so in both models the ALP couples dominantly to the first generation, with the hierarchy far more pronounced in the type-U model. This feature persists in most of the parameter space and underlies the qualitative differences between the two models discussed below.

As we have discussed, this family of ALPs can induce FCNC that are extremely suppressed in the SM. Therefore, they can be probed in flavor-physics experiments as well as the astrophysical and cosmological observations. In this section we describe each class of constraint in turn, before combining them into the exclusion plots of figure~\ref{fig:benchmarks} and discussing the implications for the two dark matter production mechanisms introduced in section~\ref{sec:relic}. 

\textit{Photon coupling and X-ray bounds.}---Tree-level ALP--meson kinetic mixing induces an effective photon coupling $g_{\phi\gamma\gamma}\propto m_\phi^2/f$ (eqs.~\eqref{eq:ALPphoton_typeD} and \eqref{eq:ALPphoton_typeU}). For the Type-D model, the coupling matrix $C_{ij}^{(D)}$ entering eq.~\eqref{eq:ALPphoton_typeD} is fixed by the mixing parameters $U_{R,4i}^{(D)}$ via eq.~\eqref{eq:CDij}, which are in turn determined by the physical quark masses, the CKM matrix, and four additional real parameters explained in section~\ref{sec:model}. The scatter plot in figure~\ref{fig:scan} reveals the hierarchy $|U_{R,43}^{(D)}| \lesssim |U_{R,42}^{(D)}| \lesssim |U_{R,41}^{(D)}|$. Evaluating eq.~\eqref{eq:ALPphoton_typeD} numerically at the benchmark eq.~\eqref{eq:benchmark_couplings} gives
\begin{equation}
g_{\phi\gamma\gamma} 
\simeq 
-6\times10^{-22}~\mathrm{GeV}^{-1} \times
\left(\frac{m_\phi}{10~\mathrm{keV}}\right)^{\!2} \times
\left(\frac{f}{10^{10}~\mathrm{GeV}}\right)^{\!-1},
\label{eq:gphigamma_D}
\end{equation}
For the Type-U model, $C_{ij}^{(U)}$ is fixed analogously via eq.~\eqref{eq:CUij}, with a stronger hierarchy $|U_{R,43}^{(U)}| \ll |U_{R,42}^{(U)}| \ll |U_{R,41}^{(U)}|$ reflecting the more severe mass hierarchy among up-type quarks. This causes $U_{R,41}^{(U)} \approx 1$ to saturate the sum rule, so $g_{\phi\gamma\gamma}$ in the Type-U model is effectively independent of $b$ and $\gamma$, with numerical value similar to eq.~\eqref{eq:gphigamma_D}. The resulting ALP lifetime,
\begin{equation}
\tau_\phi 
= \frac{64\pi}{g_{\phi\gamma\gamma}^2\,m_\phi^3}
\simeq
10^{28}~\mathrm{yr} \times 
\left(\frac{m_\phi}{10~\mathrm{keV}}\right)^{\!-7} \times
\left(\frac{f}{10^{10}~\mathrm{GeV}}\right)^{\!2},
\label{eq:lifetime}
\end{equation}
is of the same order for both models and vastly exceeds the age of the universe throughout the parameter space of interest, rendering the Nelson--Barr ALP cosmologically stable. Consequently, BBN, CMB, and $N_\mathrm{eff}$ bounds are inoperative, and stellar cooling constraints from horizontal-branch stars and white dwarfs are negligible. The only operative photon-coupling constraint comes from X-ray searches (e.g., Chandra, XMM, NuSTAR, and INTEGRAL~\cite{Panci:2022wlc}) for the quasi-monochromatic line $\phi\to\gamma\gamma$ from relic dark matter~\cite{AxionLimits}:
since $g_{\phi\gamma\gamma} \propto m_\phi^2/f$, the bound $g_{\phi\gamma\gamma} < g_\mathrm{max}(m_\phi)$ maps onto an excluded region $f\propto m_\phi^2$ at masses above $\sim 16~\mathrm{keV}$ (orange region in figures~\ref{fig:benchmarks}--\ref{fig:margins}).

\textit{SN1987A nucleon bremsstrahlung.}---Quark couplings induce ALP couplings to nucleons, enabling efficient ALP production in hot stellar plasmas~\cite{Raffelt:1996wa}. The derivative quark interactions of the Nelson--Barr ALP are matched to the nucleon sector through the non-relativistic nucleon effective Lagrangian~\cite{Badziak:2023fsc}, valid for ALP masses and relevant energies below the QCD mass gap $\Delta_{\rm QCD}\approx100$~MeV,
\begin{equation}
    \mathcal{L}_{aN}
    =
    \bar{N}v^\mu\partial_\mu N
    +\frac{\partial_\mu\phi}{f}
    \frac{C_u-C_d}{2}\,\Delta_{u-d}\,
    \bar{N}S^\mu\sigma^3 N
    +\frac{\partial_\mu\phi}{f}
    \!\left[
      \frac{C_u+C_d}{2}\,\Delta_{u+d}
      +\!\sum_{q=s,c,b,t}\!C_q\Delta_q
    \right]
    \bar{N}S^\mu N,
    \label{eq:LaN}
\end{equation}
where $N=(p,n)^T$ is the nucleon isospin doublet, $v^\mu$ its four-velocity, $2S^\mu\equiv\gamma^\mu\gamma^5$ the spin operator, and $\Delta_{u\pm d}\equiv\Delta u\pm\Delta d$. The spin fractions $\Delta q$ are extracted from lattice QCD and low-energy experiments; using the recent analysis of ref.~\cite{Badziak:2023fsc} and neglecting heavy-quark contributions, the proton and neutron couplings read
\begin{align}
    C_p &\approx 0.82\,C_u - 0.45\,C_d - 0.052\,C_s,
    \nonumber\\
    C_n &\approx 0.82\,C_d - 0.45\,C_u - 0.052\,C_s.
    \label{eq:nucleon_match}
\end{align}
Because the NB coupling is purely right-handed, $\bar{q}_R\gamma^\mu q_R=\tfrac{1}{2}(\bar{q}\gamma^\mu q +\bar{q}\gamma^\mu\gamma^5 q)$, and only the axial part contributes to eq.~\eqref{eq:nucleon_match}, so the effective Wilson coefficient is $C_q\to C^{\rm NB}_{qq}$ for each diagonal quark entry. Substituting $C_u=0$, $C_d=C_{dd}$, $C_s=C_{ss}$ for the type-D model yields,
\begin{align}
    C_p^{(D)} &= -
                0.45\,C_{dd}^{(D)}
                - 0.052\,C_{ss}^{(D)},
    \nonumber\\
    C_n^{(D)} &= 0.82\, C_{dd}^{(D)}
                     -0.052\, C_{ss}^{(D)},
    \label{eq:CpCnD}
\end{align}
and analogously $C_p^{(U)}=+0.82\,C_{uu}^{(U)}$, $C_n^{(U)}=-0.45\,C_{uu}^{(U)}$ for the type-U model, where only the $u$-quark entry contributes at leading order. The duration of the neutrino burst from SN1987A constrains the axion luminosity of the proto-neutron star via the Raffelt criterion~\cite{Carenza:2019pxu},
\begin{equation}
    \bigl(0.61\,g_{ap}^2+g_{an}^2+0.53\,g_{ap}g_{an}\bigr)
    \,\mathcal{F}\!\left(\frac{m_\phi}{T_\mathrm{SN}}\right)
    < 8.26\times10^{-19},
    \label{eq:SN_bound}
\end{equation}
where $g_{ai}\equiv C_i\,m_N/f$ with $i={n,p}$ and $T_\mathrm{SN}\simeq30$~MeV is the core temperature. The function
\begin{equation}
    \mathcal{F}(x)=\tfrac{1}{2}\,x^2 K_2(x),
    \qquad x\equiv\frac{m_\phi}{T_\mathrm{SN}},
    \label{eq:F_supp}
\end{equation}
captures the Boltzmann suppression of ALP emission from a thermal plasma: $\mathcal{F}(0)=1$ in the massless limit, while $\mathcal{F}(x)\to(\pi/2)^{1/2}x^{3/2}e^{-x}/2$ exponentially suppresses production for $m_\phi\gg T_\mathrm{SN}$~\cite{Rrapaj:2015wgs, Lella:2022uwi}. The resulting excluded band in the $(m_\phi,f)$ plane is flat for $m_\phi\ll T_\mathrm{SN}$ and retreats toward smaller $f$ for heavier ALPs, with the ChPT description itself breaking down at $m_\phi\gtrsim\Delta_{\rm QCD}$. In the equal-coupling limit $C_p\approx C_n\equiv C_N$, eq.~\eqref{eq:SN_bound} reduces to
\begin{equation}
    \frac{f}{C_N}
    \gtrsim
    1.5\times10^9\;\text{GeV}
    \times\left[\mathcal{F}\!\left(\frac{m_\phi}{T_\mathrm{SN}}\right)\right]^{1/2},
    \label{eq:SN_simple}
\end{equation}
which provides a useful order-of-magnitude estimate; the actual bound on the NB models is evaluated from eq.~\eqref{eq:SN_bound} directly using eq.~\eqref{eq:CpCnD}. Comparable constraints can be derived from neutron star cooling rates~\cite{Buschmann:2021juv}, but we use the SN1987A bound as the more conservative limit.

\textit{Red giant and white dwarf bounds.}---Cooling constraints from red giant (RG) and white dwarf (WD) stars bound the ALP--electron coupling $c_e$, generated radiatively even when $c_e(\Lambda)=0$ through electroweak loops~\cite{Bauer:2017ris}. At one loop, the dominant contribution is
\begin{equation}
    \frac{dc_e}{d\ln\mu}
    \simeq
    \frac{3y_t^2}{16\pi^2}\,c_t + \cdots,
\end{equation}
so the induced coupling is parametrically $c_e(m_e)\sim (3y_t^2/16\pi^2)\,c_t\,\ln(\Lambda/m_t)$, with subdominant contributions from lighter quarks suppressed by $y_q^2\ll y_t^2$. In the type-D model the ALP couples exclusively to right-handed down-type quarks, so $c_t=0$ identically at the UV scale, and the one-loop induced $c_e$ is suppressed by $y_b^2$. RG and WD bounds are therefore completely negligible in that model. In the type-U model, the ALP does couple to up-type quarks, but the Nelson--Barr charge matrix is strongly hierarchical. The diagonal coupling entries $C^{(U)}_{qq}=|U^{(U)}_{R,4q}|^2$ follow
\begin{equation}
    C^{(U)}_{uu} : C^{(U)}_{cc} : C^{(U)}_{tt}
    \;\approx\;
    1 : \mathcal{O}(10^{-5}) : \mathcal{O}(10^{-7}),
\end{equation}
as read directly from the scatter plot of $U_R^{(U)}$ shown in figure~\ref{fig:scan}. The coupling is almost entirely carried by the $u$ quark, precisely because the Nelson--Barr mechanism generates the right-handed rotation through the ratio $B_q/M_{\rm CP}$, which is suppressed by the up-quark Yukawa for the third generation. Consequently, despite $y_t^2\sim1$, the radiatively induced electron coupling is negligible for any reasonable UV scale. The dominant coupling $C^{(U)}_{uu}\approx1$ contributes through $y_u^2\sim10^{-10}$ and is even more suppressed. Red giant and white dwarf bounds are therefore inoperative in the type-U model as well.

\textit{Structure formation and the WDM bound.}---ALPs produced by freeze-in are born with a momentum of order the parent-quark mass and free-stream after production, suppressing primordial density fluctuations on scales below the free-streaming length ($\lambda_{\rm FS}$). This free-streaming leaves characteristic imprints on the matter power spectrum that can be probed through the Lyman-$\alpha$ forest: the absorption features in spectra of distant quasars encode the small-scale matter distribution and constrain any warm component of the dark matter~\cite{Irsic:2017ixq}. Recasting these Ly-$\alpha$ limits for the freeze-in production mechanism requires computing the full ALP velocity distribution at injection, which accounts for the shape of the phase-space population rather than just its mean momentum~\cite{DEramo:2020gpr}. The resulting ``warmness'' bound on the ALP mass takes the form~\cite{Aghaie:2024jkj}
\begin{equation}
    m_\phi
    \gtrsim
    0.01\;\text{MeV} \times 
    \left(\frac{m_{\rm WDM}}{3.5\;\text{keV}}\right)^{\!4/3} \times
    \left(\frac{70}{g_*(T_{\rm prod})}\right)^{\!1/3},
    \label{eq:WDM_bound}
\end{equation}
where $m_{\rm WDM}\approx3.5$~keV and $5.3$~keV correspond to the conservative and stringent Ly-$\alpha$ limits, respectively~\cite{DEramo:2020gpr}, and $g_*(T_{\rm prod})$ is the number of relativistic degrees of freedom at the production temperature $T_{\rm prod}\sim m_q$. For both scenarios and at the studied benchmark points, the dominant freeze-in channel is the pion scattering, therefore, we take $g_*(T_{\rm prod} \sim m_\pi)\approx 17$ in both scenarios. Substituting into eq.~\eqref{eq:WDM_bound} gives $m_\phi\gtrsim 15.9 \, \text{keV}$, for both models shown as the vertical WDM line in figure~\ref{fig:benchmarks}. 

Moreover, the Vera C.\ Rubin Observatory will measure the Milky Way subhalo mass function down to $10^6\,M_\odot$, improving the projected thermal  WDM mass limit to $m_{\rm TR} > 18\,{\rm keV}$~\cite{Baumholzer:2020hvx}.  For a photophobic freeze-in ALP (coupling primarily to SM fermions, as in our model), this translates to $m_\phi \gtrsim 62\,{\rm keV}$~\cite{Baumholzer:2020hvx},  shown as a dashed vertical line in  figure~\ref{fig:benchmarks}. This would probe the entire freeze-in window in both models essentially. Nevertheless, it is important to note that the WDM bound applies exclusively to freeze-in production; the misalignment mechanism generates a coherently oscillating condensate whose momentum redshifts to zero by the time of matter--radiation equality, so its free-streaming length is negligible and eq.~\eqref{eq:WDM_bound} is irrelevant.

\textit{Flavor bounds.}---The off-diagonal entries of $C^{(D,U)}_{ij}$ mediate tree-level FCNC decays $M\to M'\phi$. In the type-D model, the dominant constraint comes from the constraint on the $K^+$ decay measurement in the NA62 experiment, ${\rm Br}(K^+ \to \pi^+ X) \lesssim 3\times 10^{-11}$ \cite{NA62:2025upx}.
Applying the EFT given in eq.~\eqref{eq:ALP_quark_coupling_typeD} to the formulae in ref.~\cite{MartinCamalich:2020dfe} and using the lattice form factor $F_0^{K\pi}(0)=0.9698$~\cite{Aoki:2024oxs}, we obtain
\begin{align}
    {\rm Br}(K^+ \to \pi^+ \phi) \simeq 3.3 \times 10^{-9} \times \left( \frac{f}{10^{10}~{\rm GeV}} \right)^{-2} \times \left| \frac{U_{R,41}^{(D)} U_{R,42}^{(D)*}}{0.2} \right|^{2}.
\end{align}
Thus, for the type-D model, the Nelson--Barr scale is constrained as \begin{align}
    f \gtrsim 1.0 \times 10^{11}~{\rm GeV} \times \left| \frac{U_{R,41}^{(D)} U_{R,42}^{(D)*}}{0.2} \right|.
    \label{eq:NA62fbound}
\end{align}

In the type-U model, flavor-changing invisible meson decay is strongly suppressed because of a large hierarchy in $U_{R,4i}^{(U)}$. For example, we obtain ${\rm Br}(D^+ \to \pi^+ \phi) $ as
\begin{align}
    {\rm Br}(D^+ \to \pi^+ \phi) \simeq 1.9 \times 10^{-14} \times \left( \frac{f}{10^{10}~{\rm GeV}} \right)^{-2} \times \left| \frac{U_{R,41}^{(U)} U_{R,42}^{(U)*}}{0.01} \right|^{2}.
\end{align}
The current leading constraint ${\rm Br}(D^+ \to \pi^+ X) \lesssim 8.0 \times 10^{-6}$ \cite{MartinCamalich:2020dfe} is obtained from a recast of the measurement of $D^+ \to \tau^+ \nu_\tau$ at the CLEO experiment \cite{CLEO:2008ffk}.
For the type-U model, the Nelson--Barr scale is constrained as
\begin{align}
    f \gtrsim 4.9 \times 10^{5}~{\rm GeV} \times \left| \frac{U_{R,41}^{(U)} U_{R,42}^{(U)*}}{0.01} \right|.
\end{align}
which is far weaker than~\eqref{eq:NA62fbound}. Subdominant $B$-meson constraints from Belle ($B^+\to\pi^+\phi$) and Belle~II ($B^+\to K^+\phi$) probe complementary combinations of the type-D coupling matrix but do not alter the overall picture.

\textit{Implications.}---Combining the constraints above leads to sharp conclusions for both benchmark models. In the type-D model, eq.~\eqref{eq:NA62fbound} alone places the entire freeze-in production line inside the excluded region (Lower panels in figures~\ref{fig:benchmarks} and \ref{fig:margins}). More strikingly, a portion of the misalignment contours also falls within the reach of current and future precision flavour experiments: NA62 is exceptionally well suited to probe this region through the flavour-violating ALP couplings inherent to the type-D benchmark. In the type-U model on the other hand, the CLEO recast is too weak to be able to compete with the WDM bound, however, the SN1987A bound~\eqref{eq:SN_bound} can exclude a small window above $m_\phi\sim 16$~keV, but nevertheless, there is a small window in these scenarios where the ALP can be produced via Freeze-in (Upper panels in figures~\ref{fig:benchmarks} and \ref{fig:margins}). Moreover, the misalignment mechanism remains viable in both cases over the region of the misalignment contour not excluded by X-ray observations, with an open window at $m_\phi\lesssim100$~keV accessible to future high-sensitivity X-ray observatories via the $\phi\to\gamma\gamma$ line signal. In the type-U model, the misalignment contour lies beyond the reach of current precision flavour experiments; nevertheless, the Nelson--Barr ALP provides a compelling motivation for the extension and dedicated R\&D of next-generation precision experiments. Such experiments not only probe the fundamental structure of flavour in the Standard Model, but they can simultaneously probe well-motivated BSM fields like the family of anomaly-free ALPs theoretically.

\subsection{Robustness across parameter space}
\label{sec:robustness}

The constraints of section~\ref{sec:constraints} were evaluated at the single representative point of eq.~\eqref{eq:benchmark_params}. Here we promote those statements to the full parameter space of both models. The outcome is qualitatively different in the two realizations, and in both cases sharp: in the type-D model the freeze-in contour lies inside the excluded region across the overwhelming majority of parameter space, evading exclusion only in finely tuned corners; in the type-U model a viable freeze-in window survives all current constraints at \emph{every} scanned point, and will be conclusively probed by forthcoming observations. We now explain why these conclusions are, respectively, generic and universal, rather than artifacts of the chosen benchmark.

\paragraph{Factorization of the constraints.}
Once the quark masses and CKM parameters are fixed, the coupling matrix $C_{ij}=U^{*}_{R,4i}U_{R,4j}$ depends only on the four model parameters $(b,\gamma,\beta_2,\beta_3)$ of section~\ref{sec:model}. Every constraint then factorizes into a coupling scalar times a universal function of the ALP mass,
\begin{equation}
	f \;>\; f(C)\, g(m_\phi)\,,
	\label{eq:factorization}
\end{equation}
with $g(m_\phi)\sim{\rm const.}$ for the precision (laboratory and stellar) bounds and $g(m_\phi)\propto m_\phi^{3/2}$ for the X-ray line searches. The coupling factor $f(C)$ depends on a single combination of the $C_{ij}$: in the type-U model $f(C)\propto C_{uu}$ (SN1987A, X-ray), $\propto|C_{uc}|$ ($D^{+}\to\pi^{+}\phi$), $\propto C_{tt}$ (white-dwarf cooling); in the type-D model $f(C)\propto|C_{sd}|,\,|C_{bd}|,\,|C_{bs}|$ (for $K^{+}\to\pi^{+}\phi$, $B^{+}\to\pi^{+}\phi$, $B^{+}\to K^{+}\phi$ respectively), while the SN1987A bound scales with $\sqrt{X(C_{dd},C_{ss})}$ of eq.~\eqref{eq:SN_bound} and the photon-coupling bounds with the combination in eq.~\eqref{eq:gphigamma_D}. Likewise, since every freeze-in rate scales as $|C|^{2}/f^{2}$, the contour reproducing the observed relic abundance obeys
\begin{equation}
	f_{\rm FI}(m_\phi)
	= \left[\frac{m_\phi\, s_0}{\rho_c\,\Omega_{\rm DM}h^{2}}
	\sum_{\rm ch} |C_{\rm ch}|^{2}\, Y^{(1)}_{\rm ch}\right]^{1/2},
	\label{eq:fFI}
\end{equation}
with $Y^{(1)}_{\rm ch}$ the yield of each channel at unit coupling, computed once and for all in section~\ref{sec:direct_freezein}. Consequently, the Boltzmann equations are solved a single time per channel, and evaluating the full set of constraints at any parameter point reduces to algebra: the parameter space can be scanned efficiently at high resolution. We draw $N=\Nscan$ points uniformly over $b\in(\bMin,\bMax)$\footnote{The range of $b$ follows from the unitarity of $U_R$ and the reality of the mass-matrix square roots, which force $a,b\in(0,1)$ with $a^2+b^2<1$; see appendix~\ref{app:mu a b}.} and $\gamma,~\beta_2,~\beta_3\in[0,2\pi)$.

\begin{figure}[p]
  \centering
  \includegraphics[width=\textwidth]{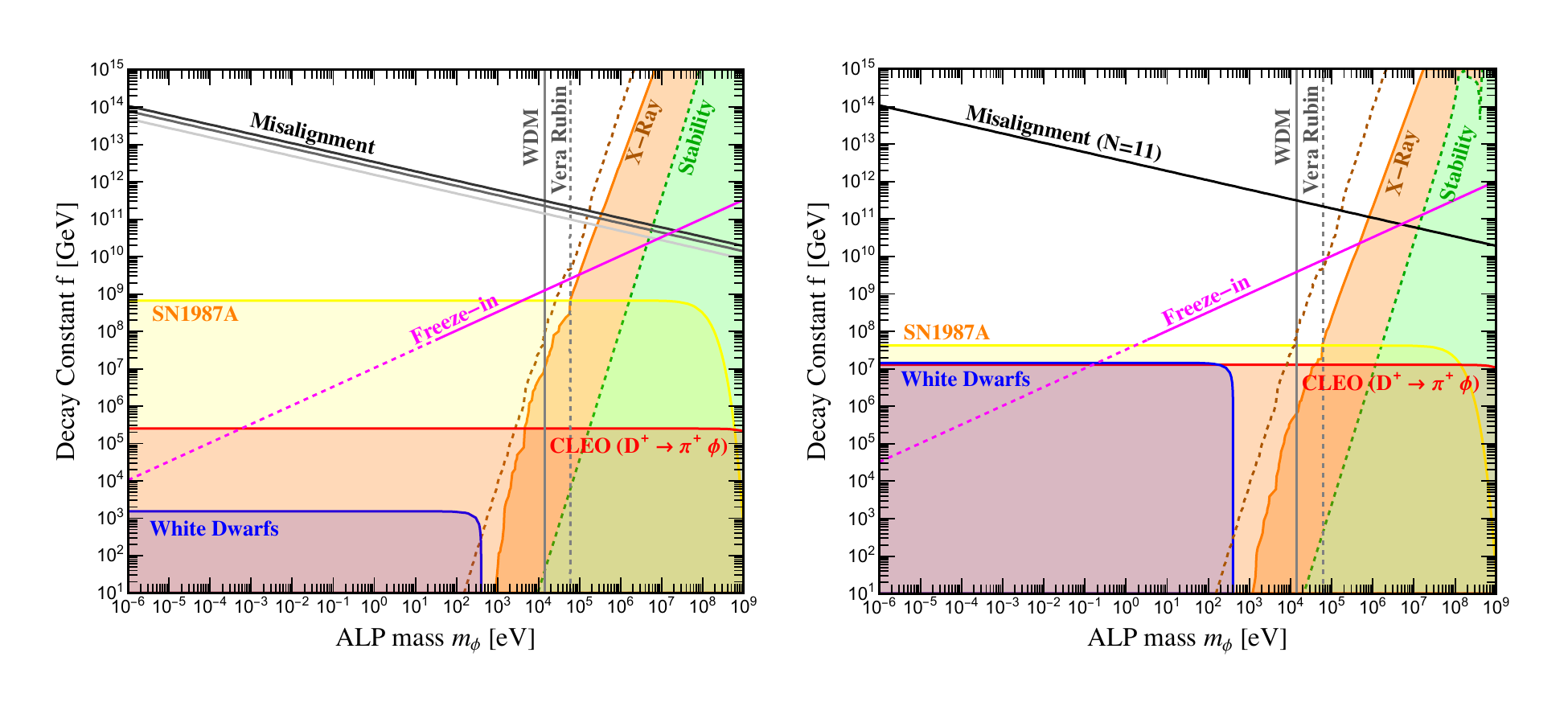}\\[4pt]
  \includegraphics[width=\textwidth]{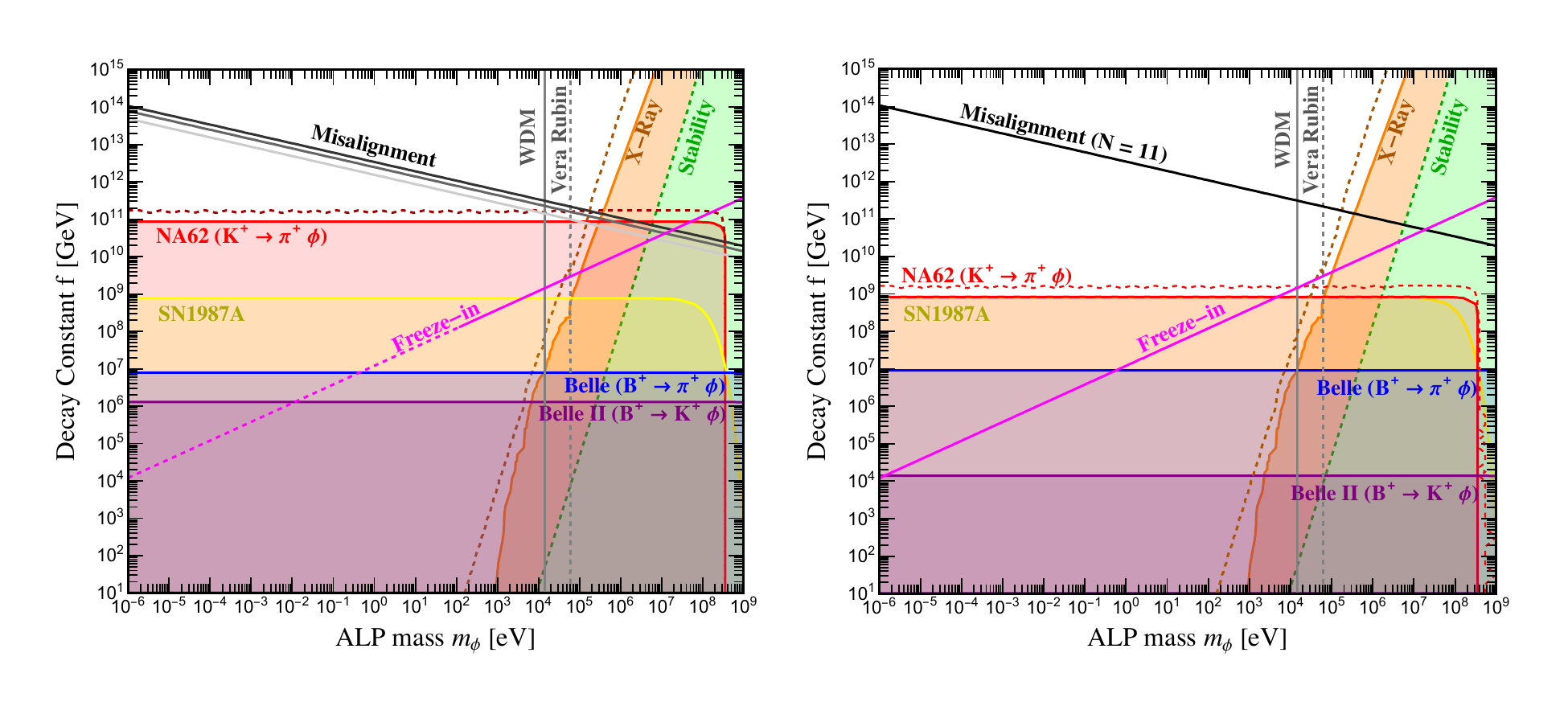}
  \caption{Constraints on the type-U (upper panels) and type-D (lower panels) models, evaluated at the representative benchmark \eqref{eq:benchmark_couplings} (left) and at the adversarial benchmark minimizing the margin $\mathcal{R}$ discussed in section \ref{sec:robustness} (right). Shaded regions are excluded: SN1987A~\cite{Carenza:2019pxu} (light yellow), the leading flavor bounds $D^{+}\to\pi^{+}\phi$ from CLEO~\cite{MartinCamalich:2020dfe, CLEO:2008ffk} (type-U) and $K^{+}\to\pi^{+}\phi$ from NA62~\cite{NA62:2025upx} (type-D) in red, the requirement that $\tau_\phi$ exceeds the age of the Universe (green), and X-ray line searches (orange)~\cite{Panci:2022wlc}, whose projected future sensitivity is shown as the dashed orange contour. The vertical gray line marks the warm-dark-matter bound~\cite{DEramo:2020gpr}, applicable only to freeze-in production; the dashed vertical line shows the projected improvement from the Vera C.\ Rubin Observatory~\cite{Baumholzer:2020hvx}. The solid magenta line reproduces the observed relic abundance via freeze-in (dashed for $f \lesssim f_{\rm th}$, below which the ALP thermalizes with the SM bath and freeze-in is no longer valid), and the gray curves via misalignment for $Z_N$ with $N = 5,\,8,\,11$ (light to dark). Here $\theta_0 \equiv \phi_{\rm osc}/f$; since the minima sit at $\phi = 2\pi k f/N$, the displacement from the nearest minimum obeys $|\theta_0| \le \pi/N$, and for each curve we take the root-mean-square value of a uniformly distributed initial angle, $\theta_0 = \pi/(N\sqrt{3})$, so that no tuning of the initial condition is assumed. Because $f_{\rm mis} \propto 1/\theta_0$, the curves scale linearly with $N$. In the type-D model, the NA62 bound buries the freeze-in contour for most of the parameter space but the freeze-in window survives in a small portion of scanned points that is most prominent at the adversarial point, whereas in the type-U model the freeze-in window remains viable for the vast majority of scanned points. This window is between the warm-dark-matter and X-ray bounds and it will be probed complementarily by future galaxy surveys and X-ray observations.
  }
  \label{fig:benchmarks}
\end{figure}

\paragraph{Exclusion margin.}
Freeze-in dark matter populates the single contour $f=f_{\rm FI}(m_\phi)$ rather than a region of the $(m_\phi,f)$ plane. For each parameter point we therefore define the exclusion margin
\begin{equation}
	\mathcal{R}
	\;\equiv\;
	\min_{m_\phi \,\in\, [m_{\rm WDM},\,10^{8}\,{\rm eV}]}
	\frac{f_{\rm excl}(m_\phi)}{f_{\rm FI}(m_\phi)}\,,
	\qquad
	f_{\rm excl}(m_\phi)=\max_{a}\, f_{a}(m_\phi)\,,
	\label{eq:margin}
\end{equation}
with $a$ running over the current bounds and the mass window starting at the Lyman-$\alpha$ limit $m_{\rm WDM}$, below which freeze-in is excluded by structure formation for any $f$. By construction $\mathcal{R}>1$ means the entire freeze-in contour lies inside the excluded region, while $\mathcal{R}<1$ signals a surviving mass window in which the contour rises above every current bound.
In addition to the representative benchmark point \eqref{eq:benchmark_couplings}, We also take the \emph{adversarial} benchmark point in which ${\cal R}$ is minimized among the scanned points. This benchmark point is shown in figure \ref{fig:scan} and the right panels of figure \ref{fig:benchmarks}.

\paragraph{Why the conclusions are robust: a universal contour against mobile bounds.}
The central observation is a hierarchy of sensitivities. As $(b,\gamma,\beta_2,\beta_3)$ range over the scan, the flavor bounds move by orders of magnitude in $f$, tracking the flavor-violating couplings $|C_{sd}|$ (type-D) and $|C_{uc}|$ (type-U) down to the $10^{-5}$--$10^{-3}$ level, while the freeze-in contour barely moves (appendix~\ref{app:scan_ranges} collects the variation factors). The reason is structural: the contour is pinned from below by pion scattering $\pi^{+}\pi^{-}\to\pi^{0}\phi$, whose yield scales as $(C_{dd}-C_{uu})^2$, and the Nelson--Barr hierarchy forces the relevant diagonal coupling to unity in most of the parameter space---$C_{dd}\simeq1$ in the type-D model, $C_{uu}=1-|U_{R,42}|^{2}-|U_{R,43}|^{2}-|U_{R,44}|^{2}\simeq1$ in the type-U model. The contour is therefore a nearly fixed curve while the bounds sweep across it. Whether freeze-in survives is then decided not by how the contour moves, but by whether the \emph{operative} bound is coupled to the \emph{same} unsuppressed diagonal coupling that pins the contour. This single criterion separates the two models, and underlies both results below.

\paragraph{Results: type-D.}
Here the flavor bounds are strong. At the benchmark of eq.~\eqref{eq:benchmark_params} we find $\mathcal{R}=\RbenchD$, deep inside the excluded region, with $f_{\rm FI}(1~{\rm MeV})=\fFIbenchD$ and a mesonic yield share of $\DmesShare$. Raising the contour above the bounds requires enhancing the flavor-violating channels $b\to d\phi$ and $K\to\pi\phi$, but the couplings that control them, $|C_{bd}|$ and $|C_{sd}|$, simultaneously set the heights of the $B^{+}\to\pi^{+}\phi$ and $K^{+}\to\pi^{+}\phi$ bounds, which strengthen in step; and the SN1987A floor, fixed by $X(C_{dd},C_{ss})$ with $C_{dd}\simeq1$, tracks the pinned contour from below.
Since $C_{ij}$ is determined as $C_{ij} = U_{R,4i}^* U_{R,4j}$., evading exclusion therefore requires a strong suppression of $U_{R,42}$ while keeping $U_{R,41} \simeq 1$.
Such points do exist but are rare and tuned: the most freeze-in-favorable point of our $N=\Nscan$ scan reaches $\mathcal{R}_{\rm min}=\RminD$, and the minimum continues to decrease as the sampling is refined (we find $\mathcal{R}_{\rm min}\simeq\RminDbig$ at $N=\NscanBig$), meaning that a portion of freeze-in window survives for about $0.025\%$ of the scanned points. We therefore do not claim exclusion at literally every point; the robust statement is that freeze-in production of the type-D Nelson--Barr ALP is excluded across all of the parameter space except a finely tuned subset.

\paragraph{Results: type-U.}
The type-U model behaves in the opposite way, and here the conclusion is almost universal. Freeze-in is dominated by pion scattering with yield $\propto C_{uu}^{2}$ and $C_{uu}\simeq1$ in most of the parameter space, so the contour is nearly fixed. Crucially, the operative bounds share this coupling: both the SN1987A floor and the X-ray limits scale with $C_{uu}$, and both lie \emph{below} the contour in a finite mass range. The flavor probe $D^{+}\to\pi^{+}\phi$ (CLEO), scaling with the strongly suppressed $|C_{uc}|$, is weaker than its kaon counterpart by orders of magnitude in branching-ratio reach and never approaches the contour, indeed, this bound alone sweeps by $\CLEOShift$ in $f$ across the scan without ever crossing it. We consequently find $\mathcal{R}<1$ for \emph{every} scanned point: the margin ranges from $\mathcal{R}=\RadvU$ at the most freeze-in-favorable (weakest-constraint) point to a maximum of $\mathcal{R}=\RmaxU$ at the most constrained point. The fact that even the most constrained type-U point remains below unity is what makes the surviving window a parameter-independent prediction rather than a tuned feature. The window, $m_\phi\in\survMassU$, is bounded from below by the Lyman-$\alpha$ limit and from above by the X-ray bounds (figure~\ref{fig:benchmarks}, upper right).

\begin{figure}[p]
  \centering
  \includegraphics[width=1.03\textwidth]{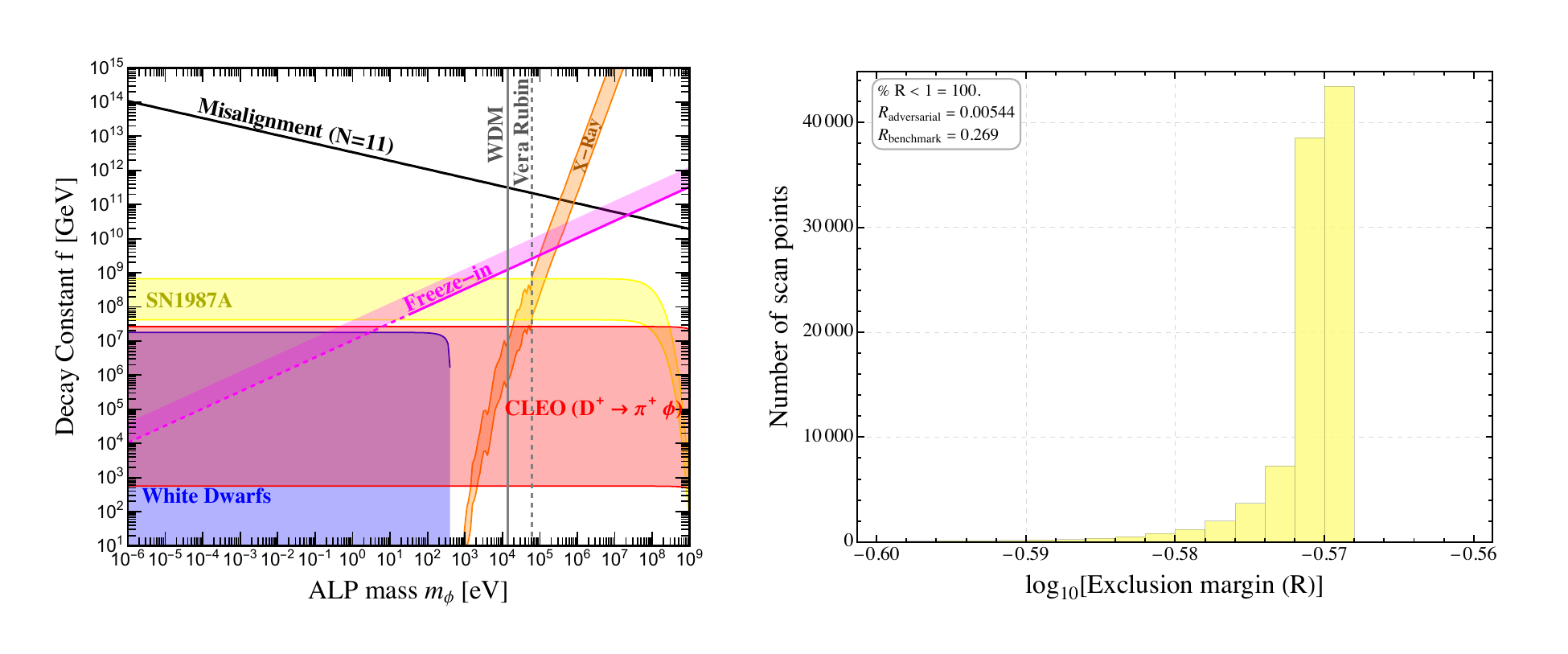}\\[4pt]
  \includegraphics[width=1.03\textwidth]{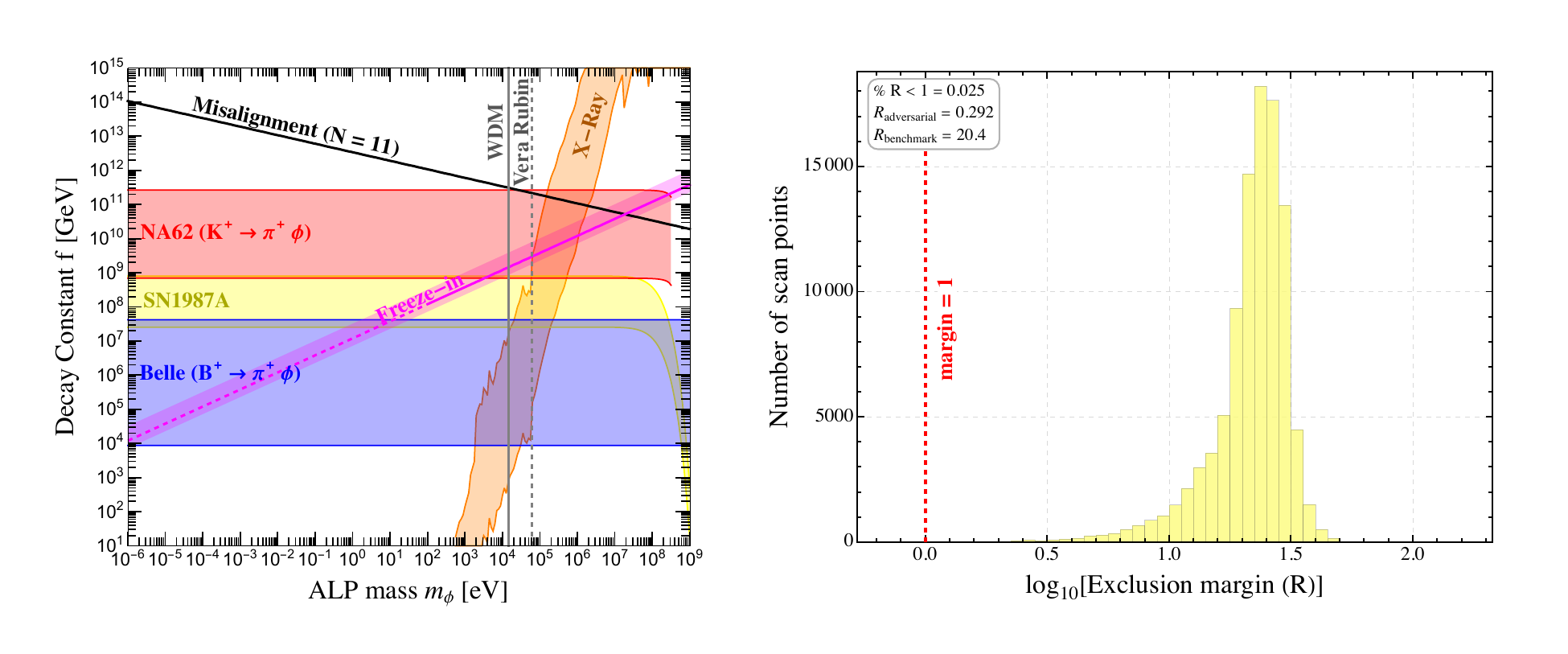}
    \caption{Benchmark dependence of the constraints for the type-U (upper) and type-D (lower) models. Left: instead of exclusion regions, the shaded bands show the full \emph{variation} of each constraint boundary and of the freeze-in contour (magenta) across the parameter scan; only the dominant constraints are displayed. Right: distribution of the exclusion margin $\mathcal{R}$ over the scan. In the type-D model $\mathcal{R}>1$ for more than $99.9\%$ of the scanned points, failing only in a finely tuned corner: the freeze-in contour is pinned by the $\pi\pi\to\pi\phi$ channels ($\propto C_{dd}\simeq1$, $\DsharePiPi$ of the abundance) and the SN1987A floor tracks it, while raising the contour above the bounds requires enhancing the flavor-violating channels, $K \to \pi \phi $ decay in particular, whose coupling $|C_{sd}|$ determines the NA62 bound, so exclusion can be evaded only in tuned corners. In the type-U model, by contrast, the flavor bound from $D^{+}\to\pi^{+}\phi$ is much weaker, while the operative SN1987A and X-ray bounds scale with the same $C_{uu}\simeq1$ that pins the contour and fall below it in a finite mass range: $\mathcal{R}<1$ for \emph{every} scanned point, and the freeze-in window is viable for $m_\phi\sim\survMassLoU$--$\survMassHiU$. This surviving region will be probed on two complementary fronts: the Vera~C.\ Rubin Observatory will strengthen the warm-dark-matter bound (dashed vertical line), and future X-ray missions will push the orange boundary downward, together covering the entire freeze-in window.}
  \label{fig:margins}
\end{figure}

Two remarks place this window in perspective. First, it lies entirely within the reach of forthcoming probes attacking it from both sides: the Vera~C.\ Rubin Observatory measurement of the subhalo mass function will push the warm-dark-matter bound to $m_\phi\lesssim\RubinWDM$ for a photophobic freeze-in ALP (dashed vertical line in figures~\ref{fig:benchmarks} and \ref{fig:margins}), while future X-ray missions will lower the photon-coupling boundary from above, together covering the window, so that type-U freeze-in will be conclusively tested. Second, the width of the window is sensitive to the SN1987A bound, which carries order-of-magnitude systematics: a modest strengthening would narrow it from below. The type-D conclusion, by contrast, is insensitive to such systematics, being driven by NA62 alone. The misalignment prediction is unaffected by any of the above: at the untuned angle $\theta_0=\pi/(N\sqrt{3})$ the contour shifts rigidly as $f_{\rm mis}\propto N$, so varying $N$ over the range shown
displaces it by only a factor of two which does not affect the allowed window.

\section{Conclusion}
In this paper, we have studied Nelson--Barr models with a $Z_N$ symmetry as a solution to the strong CP problem. As discussed in section~\ref{sec:EFT}, Nelson--Barr models with $N\geq 5$ have an accidental $U(1)$ symmetry. This $U(1)$ symmetry is spontaneously broken by the vacuum expectation value of the Nelson--Barr scalar field $\Phi$ and explicitly broken by higher-dimensional operators. Consequently, an ALP with a small mass arises naturally in this class of models. This ALP is anomaly-free, and its couplings to photons and gluons are suppressed by the square of the ALP mass. The couplings between the ALP and the SM quarks are determined by the matrix $U_R$ introduced in section \ref{sec:model}, and the ALP tends to couple most strongly to first-generation quarks.

In section \ref{sec:anomalyfreeALP}, we investigated the possibility that the ALP constitutes dark matter.
As shown in figure \ref{fig:benchmarks}, for the representative benchmark point in the type-D model, the parameter region in which freeze-in produces the dominant component of the observed dark-matter abundance is excluded by the NA62 constraint in the Type-D model.
For the type-U model, the SN1987A observation gives a severe constraint, and only a very limited mass range $m_\phi \sim [16 - 100]~{\rm keV}$ is available.
We also scanned the parameter spaces of both the Type-U and Type-D models. As shown in figure \ref{fig:margins}, in the Type-D model, almost the entire parameter region in which freeze-in accounts for the dominant component of dark matter is excluded, evading exclusion only in finely tuned corners. In the Type-U model, an allowed parameter region remains for freeze-in ALP dark matter at \emph{every} scanned point, making it a robust prediction of the model rather than a tuned feature; this region can be probed by future observations with the Vera C.~Rubin Observatory, which will push the warm-dark-matter bound to $m_\phi\lesssim\RubinWDM$, together with forthcoming X-ray line searches such as GECCO and THESEUS/XGIS-S.
If dark matter is instead produced predominantly through the misalignment mechanism, no severe constraints arise, and this scenario requires $f\gtrsim 10^{11}~{\rm GeV}$ and $m_\phi\lesssim 100~{\rm keV}$.
Thus, Nelson--Barr models with sufficiently large discrete symmetries provide a simple framework in which a solution to the strong CP problem is accompanied by a naturally light, anomaly-free ALP with a predictive flavor structure and viable, imminently testable dark-matter phenomenology. 

Importantly, the ALP in this scenario is not an optional add-on but an unavoidable consequence of the accidental $U(1)$ that the $Z_N$ symmetry protects: its mass, its suppressed photon and gluon couplings, and its flavor structure are all fixed by the same physics that solves the strong CP problem. This makes the ALP a genuine messenger of the Nelson--Barr mechanism.
Whereas the mechanism itself operates at the high scale $f$ and is otherwise difficult to probe directly, the detection of a light, anomaly-free, first-generation-philic ALP through the X-ray and warm-dark-matter searches above, or through the flavor transitions $K, B\to\pi\,\phi$ and $D\to\pi\,\phi$, would furnish a novel, low-energy handle on Nelson--Barr mechanism as the origin of the vanishing strong CP phase, turning an otherwise inaccessible high-scale construction into a falsifiable proposal.

\section*{Acknowledgements}
This work is supported in part by JSPS KAKENHI Grant Numbers~23K03415 (RS), 24H02236 (RS), 24H02244 (MA and RS), and 26K24541 (MA).

\appendix

\section{Parameter ranges, scan, and sensitivity of the constraints}
\label{app:scan_ranges}

This appendix supports the robustness analysis of section~\ref{sec:robustness}. We first derive the allowed ranges of the free parameters that define the scan, then tabulate the raw coupling ranges and benchmark values, and finally quantify the sensitivity of each constraint across the parameter space.

\subsection{Allowed ranges of the free parameters}
\label{app:mu a b}

We derive the allowed ranges of $\mu$, $a$, and $b$ introduced in section~\ref{sec:model}, which fix the domain of the scan. Define the matrix
\begin{align}
	A \equiv V_{d_L}\widehat M_d^2 V_{d_L}^\dagger. \label{eq:def A matrix}
\end{align}
Using eqs.~\eqref{eq:definition X} and \eqref{eq:definition Z}, its real and imaginary parts are
\begin{align}
	{\rm Re}\, A = X^2, \qquad
	{\rm Im}\, A = XZX,
\end{align}
so that $A = X(I_3 + i Z)X$, and
\begin{align}
	\det A
	= (\det X)^2 \det (I_3+iZ)
	= (\det X)^2 (1-\mu^2), \label{eq:detA 1}
\end{align}
where we used eqs.~\eqref{eq:P C mu} and \eqref{eq:a b mu}. On the other hand, from the definition \eqref{eq:def A matrix}, the determinant is manifestly positive,
\begin{align}
	\det A = \det \widehat M_d^2 = m_d^2 m_s^2 m_b^2 > 0. \label{eq:detA 2}
\end{align}
Comparing eqs.~\eqref{eq:detA 1} and \eqref{eq:detA 2} gives $0 \leq \mu < 1$; requiring a physical CP phase in the CKM matrix excludes $\mu=0$, so
\begin{align}
	0 < \mu < 1.
\end{align}
The parameters $a$ and $b$ satisfying eq.~\eqref{eq:a b mu} can be parametrized by a single real number $x$,
\begin{align}
	a = \sqrt{ \frac{\mu e^x}{1 + \mu e^x} }, \qquad
	b = \sqrt{ \frac{\mu e^{-x}}{1 + \mu e^{-x}} },
	\qquad -\infty < x < \infty,
	\label{eq:ab param}
\end{align}
from which $0 < a < 1$ and $0 < b < 1$.
As we can see that $a$ and $b$ are parametrized a single parameter $x$, they are not independent to satisfy eq.~\eqref{eq:a b mu}.
In particular, since $0< \mu < 1$, $a^2 + b^2$ is constrained as
\begin{align}
	a^2 + b^2
	= 1 - \frac{1-\mu^2}{1 + \mu e^{x} + \mu e^{-x} + \mu^2} \;<\; 1,
\end{align}
Thus, the parameter space of $a$ and $b$ is therefore the interior of the unit quarter-disk, $a,b\in(0,1)$ with $a^2+b^2<1$. Accordingly we scan $b$ over $(\bMin,\bMax)\simeq[\epsilon,1-\epsilon]$ (with $a$ fixed by eq.~\eqref{eq:a b mu} for each point) and the three phases $\gamma,\beta_2,\beta_3$ over $[0,2\pi)$. The same argument holds for the type-U model under $\widehat M_d\to\widehat M_u$.

\subsection{Scan setup and coupling ranges}
\label{app:scan_setup}

Once the quark masses and CKM parameters are fixed, each scan point reduces to the set of coupling scalars on which the constraints of section~\ref{sec:constraints} depend (eq.~\eqref{eq:factorization}). We draw $N=\Nscan$ points uniformly over the domain above. Table~\ref{tab:scanrange} gives, for each scalar, its range over the scan and its value at the benchmark of eq.~\eqref{eq:benchmark_params}, together with the freeze-in scale $f_{\rm FI}(1~{\rm MeV})$ and the exclusion margin $\mathcal{R}$ of eq.~\eqref{eq:margin}.

\begin{table}[h]
	\renewcommand{\arraystretch}{1.25}\centering\small
	\begin{tabular}{c l r@{\ \ldots\ }l r}
		\hline\hline
		& Quantity & \multicolumn{2}{c}{scan range} & benchmark \\
		\hline
		\multirow{8}{*}{Type-U}
		& $C_{uu}$ \small(SN, X-ray, $g_{\phi\gamma\gamma}$) & $6.27\times10^{-2}$ & $1.00$               & $1.00$ \\
		& $|C_{uc}|$ \small(CLEO)        & $1.05\times10^{-5}$  & $0.495$              & $4.71\times10^{-3}$ \\
		& $|C_{ut}|$                       & $2.60\times10^{-6}$  & $7.94\times10^{-2}$  & $9.51\times10^{-4}$ \\
		& $|C_{ct}|$                       & $1.98\times10^{-9}$  & $8.87\times10^{-2}$  & $4.48\times10^{-6}$ \\
		& $C_{cc}$                         & $1.11\times10^{-10}$ & $0.927$              & $2.22\times10^{-5}$ \\
		& $C_{tt}$ \small(white dwarfs)    & $6.76\times10^{-12}$ & $1.05\times10^{-2}$  & $9.04\times10^{-7}$ \\
		& $f_{\rm FI}(1\,{\rm MeV})$ [GeV] & $1.05\times10^{10}$  & $4.06\times10^{10}$  & $1.05\times10^{10}$ \\
		& $\mathcal{R}$                    & $5.44\times10^{-3}$  & $0.269$              & $0.269$ \\
		\hline
		\multirow{9}{*}{Type-D}
		& $C_{dd}$                         & $3.70\times10^{-3}$ & $1.00$              & $0.965$ \\
		& $C_{ss}$                         & $1.69\times10^{-6}$ & $0.919$             & $2.73\times10^{-2}$ \\
		& $C_{bb}$                         & $8.91\times10^{-9}$ & $0.586$             & $7.33\times10^{-3}$ \\
		& $|C_{sd}|$ \small(NA62)          & $1.30\times10^{-3}$ & $0.499$             & $0.162$ \\
		& $|C_{bd}|$ \small(Belle)         & $9.33\times10^{-5}$ & $0.449$             & $8.41\times10^{-2}$ \\
		& $|C_{bs}|$ \small(Belle II)      & $7.29\times10^{-6}$ & $0.473$             & $1.41\times10^{-2}$ \\
		& $\sqrt{X}$ \small(SN1987A)       & $2.48\times10^{-2}$ & $0.775$             & $0.746$ \\
		& $f_{\rm FI}(1\,{\rm MeV})$ [GeV] & $7.22\times10^{9}$  & $3.21\times10^{10}$ & $1.20\times10^{10}$ \\
		& $\mathcal{R}$                    & $0.292$             & $52.2$              & $20.4$ \\
		\hline\hline
	\end{tabular}
	\caption{Scan ranges and benchmark values of the coupling scalars, freeze-in scale, and exclusion margin for the type-U (upper) and type-D (lower) models ($N=\Nscan$ points). Every type-U point has $\mathcal{R}<1$, and the benchmark lies within $0.1\%$ of the largest margin encountered, so it is the most constrained point of the scan. The type-D benchmark lies deep in the excluded region, $\mathcal{R}=\RbenchD$; freeze-in survives only in a corner with $\mathcal{R}<1$, whose extent is discussed in appendix~\ref{app:sensitivity}.}
	\label{tab:scanrange}
\end{table}

\subsection{Sensitivity of the bounds across the scan}
\label{app:sensitivity}

The robustness argument of section~\ref{sec:robustness} rests on a hierarchy of sensitivities, which is read directly from the ranges above. Since $f_{\rm excl}\propto f(C)$, each bound's variation across the scan equals that of its coupling scalar, whereas the freeze-in contour is protected by the near-unit diagonal coupling $C_{dd},~C_{uu} \simeq 1$. Quantitatively, the freeze-in contour varies by only a factor $\fFIshiftD$ (type-D) and $\fFIshiftU$ (type-U) in $f$, while the flavor bounds sweep across it by orders of magnitude: the $K^{+}\to\pi^{+}\phi$ bound of NA62 shifts by $\NAsixtwoShift$ (type-D) and the $D^{+}\to\pi^{+}\phi$ bound of CLEO by $\CLEOShift$ (type-U), while the SN1987A floor shifts by $\SNshiftD$ (type-D). This is the quantitative content of the ``mobile bounds against a universal contour'' picture: whether freeze-in survives is decided by whether the operative bound is tied to the same unsuppressed diagonal coupling that pins the contour.

For reproducibility, we record the adversarial (margin-minimizing) points, which set the deepest surviving windows and are shown in the right panels of figure~\ref{fig:margins}. In the type-D model the minimum margin $\mathcal{R}=\RminD$ is reached at $(b,\gamma,\beta_2,\beta_3)= (0.0348,\,3.46,\,3.12,\,3.09)$; the corresponding coupling matrix is
\begin{align}
	|C^{(D)}_{\rm adv}| =
	\begin{pmatrix}
	0.990 & 1.56\times10^{-3} & 9.71\times10^{-2} \\
	1.56\times10^{-3} & 2.46\times10^{-6} & 1.53\times10^{-4} \\
	9.71\times10^{-2} & 1.53\times10^{-4} & 9.52\times10^{-3}
	\end{pmatrix},
\end{align}
illustrating the required simultaneous suppression of the flavor-violating entries and of $C_{ss}$ while $C_{dd}\simeq1$ is retained. In the type-U model the minimum margin $\mathcal{R}=\RadvU$ is reached at $(b,\gamma,\beta_2,\beta_3)= (0.238,\,5.69,\,6.28,\,1.19)$, corresponding to
\begin{align}
	|C^{(U)}_{\rm adv}| =
	\begin{pmatrix}
	6.27\times10^{-2} & 0.241 & 2.31\times10^{-2} \\
	0.241 & 0.927 & 8.87\times10^{-2} \\
	2.31\times10^{-2} & 8.87\times10^{-2} & 8.48\times10^{-3}
	\end{pmatrix}.
\end{align}
For the type-D model, we find that cancellation in the couplings relevant to physical bounds happens in a corner of the parameter space. We did not observe convergence of ${\cal R}_{\rm min}$, and it continues to decrease with sample size ($\mathcal{R}_{\rm min}\simeq\RminDbig$ at $N=\NscanBig$).

\section{Meson effective theory for the Nelson--Barr ALP}
\label{app:meson_eft}

This appendix derives the low-energy meson effective theory used in the freeze-in calculation of section~\ref{sec:direct_freezein} and collects all hadronic amplitudes that enter the numerical analysis. Throughout this appendix we focus on the type-D model, for which the ALP couples directly to the light down-type quarks and mesonic freeze-in production is relevant; the corresponding statements for the type-U model are summarized in appendix~\ref{app:typeU_remark}. Above the QCD confinement scale, the relevant degrees of freedom are quarks and gluons, and the ALP couples through the derivative interaction of eq.~\eqref{eq:ALP quark coupling type-D},
\begin{equation}
    \mathcal{L}_{\phi q} = \frac{\partial_\mu\phi}{f} \, C^{(D)}_{ij} \, \bar{d}_{Ri} \gamma^\mu d_{Rj},
    \qquad i,j = d,s,b,
    \label{eq:app_phi_quark}
\end{equation}
with $C^{(D)}_{ij} = U^{(D)*}_{R,4i} U^{(D)}_{R,4j}$ as defined in eq.~\eqref{eq:CDij}. Note that $C^{(D)}$ is Hermitian by construction, $C^{(D)}_{ds} = C^{(D)*}_{sd}$. In the following, we drop the superscript and write $C_{ij}\equiv C^{(D)}_{ij}$. 

Because eq.~\eqref{eq:app_phi_quark} is a flavor-dependent coupling to right-handed down-type quarks rather than an anomalous $\phi G\widetilde G$ interaction, the meson effective theory it induces differs qualitatively from that of the QCD axion. Below we derive it via chiral perturbation theory (ChPT), diagonalize the neutral sector, and list all interaction vertices and amplitudes relevant for freeze-in production. We work at leading order in the chiral expansion and at first order in $1/f$; corrections of order $f_\pi^2/f^2$ and $p^4/\Lambda_\chi^4$ (with $\Lambda_\chi \sim 4\pi f_\pi$) are neglected, and we take the isospin limit for meson masses inside amplitudes. We also assume $m_\phi \ll m_\pi$, which is the regime relevant for freeze-in; the leading effects of a finite ALP mass are commented on where they matter.

\subsection{Chiral matching}
\label{app:chiral_matching}

Our starting point is the $U(3)$ chiral Lagrangian of eq.~\eqref{eq: Chiral Lag},
\begin{equation}
    \mathcal{L}_{\chi\rm eff}
    = \frac{f_\pi^2}{4} \operatorname{Tr}\!\left[ D_\mu\Sigma \left(D^\mu\Sigma\right)^\dagger \right]
    + \frac{f_\pi^2 B_0}{2} \operatorname{Tr}\!\left[ M_q \Sigma^\dagger + \mathrm{h.c.} \right]
    - \frac{1}{2} M_0^2 \eta_0^2 ,
    \label{eq:app_LO_chiral}
\end{equation}
with $M_q = \operatorname{diag}(m_u, m_d, m_s)$, $f_\pi = 93$~MeV, and
\begin{equation}
    \Sigma = \exp\!\left[ i\frac{\sqrt{2}}{f_\pi}\,\Pi \right],
    \qquad
    \Pi = \begin{pmatrix}
            \dfrac{\pi^0}{\sqrt{2}} + \dfrac{\eta_8}{\sqrt{6}} + \dfrac{\eta_0}{\sqrt{3}} & \pi^+ & K^+ \\[3mm]
            \pi^- & -\dfrac{\pi^0}{\sqrt{2}} + \dfrac{\eta_8}{\sqrt{6}} + \dfrac{\eta_0}{\sqrt{3}} & K^0 \\[3mm]
            K^- & \bar{K}^0 & -\dfrac{2\eta_8}{\sqrt{6}} + \dfrac{\eta_0}{\sqrt{3}}
          \end{pmatrix},
    \label{eq:app_meson_matrix}
\end{equation}
as in the definition following eq.~\eqref{eq: Chiral Lag}. The flavor-singlet $\eta_0$ is retained together with the anomaly-induced mass term $-\tfrac12 M_0^2\eta_0^2$, so that the physical $\eta$ and $\eta'$ can be defined within the effective theory; this is the standard large-$N_c$ extension of $SU(3)$ ChPT. Since the singlet component of $\Pi$ is proportional to the identity matrix, it drops out of all commutators below and enters only through the kinetic-mixing term \eqref{eq:app_L1}.

The ALP enters as a right-handed external source. Writing eq.~\eqref{eq:app_phi_quark} as $\bar q_R \gamma^\mu r_\mu q_R$ in the $(u,d,s)$ basis fixes
\begin{equation}
    r_\mu = \frac{\partial_\mu\phi}{f}\, C', \qquad \ell_\mu = 0,
    \qquad
    C' = \begin{pmatrix} 0 & 0 & 0 \\ 0 & C_{dd} & C_{ds} \\ 0 & C_{sd} & C_{ss} \end{pmatrix},
    \label{eq:app_spurion}
\end{equation}
as in the definition of $C'^{(D)}$ in the text; the vanishing first row and column reflect the absence of an up-quark coupling in the type-D model. Note that no additional factor of $1/2$ appears in this identification: in the external-source formalism the right-handed source couples as $\mathcal{L}_{\rm ext} = \bar q_R\gamma^\mu r_\mu q_R$, so the coefficient of $r_\mu$ is read off directly from eq.~\eqref{eq:app_phi_quark}. (The familiar factor $q_R\gamma^\mu q_R = \tfrac12\,\bar q\gamma^\mu(1+\gamma_5)q$ is instead relevant when matching to \emph{vector}-current matrix elements, and is automatically reproduced by the results below; see the remark after eq.~\eqref{eq:app_GammaKpi}.) The covariant derivative is that of eq.~\eqref{eq:ALP meson interaction}, and expanding the kinetic term of eq.~\eqref{eq:app_LO_chiral} to first order in $\phi$ yields the master interaction of eq.~\eqref{eq:ALP meson interaction},
\begin{equation}
    \mathcal{L}_{\phi\chi}
    = -\frac{i f_\pi^2}{4f}\, \partial_\mu\phi \, \operatorname{Tr}\!\left[ (\partial^\mu\Sigma)\, C' \,\Sigma^\dagger \right] + \mathrm{h.c.}
    = -\frac{i f_\pi^2}{2f}\, \partial_\mu\phi \, \operatorname{Tr}\!\left[ C'\, \Sigma^\dagger \partial^\mu\Sigma \right],
    \label{eq:app_master}
\end{equation}
where the second form follows because $\operatorname{Tr}[C'\Sigma^\dagger\partial\Sigma]$ is anti-Hermitian for Hermitian $C'$, so the two terms of the first form are equal. All meson--ALP interactions used in this work descend from eq.~\eqref{eq:app_master}.

\subsection{Expansion in meson fields}
\label{app:expansion}

Using
\begin{equation}
    \Sigma^\dagger \partial_\mu\Sigma
    = \frac{i\sqrt{2}}{f_\pi}\, \partial_\mu\Pi
    + \frac{1}{f_\pi^2}\, [\Pi, \partial_\mu\Pi]
    - \frac{i\sqrt{2}}{3f_\pi^3}\, [\Pi, [\Pi, \partial_\mu\Pi]]
    + \mathcal{O}(\Pi^4),
    \label{eq:app_current_expansion}
\end{equation}
eq.~\eqref{eq:app_master} decomposes as $\mathcal{L}_{\phi\chi} = \mathcal{L}^{(1)} + \mathcal{L}^{(2)} + \mathcal{L}^{(3)} + \cdots$ with
\begin{align}
    \mathcal{L}^{(1)} &= \frac{f_\pi}{\sqrt{2} f}\, \partial_\mu\phi \, \operatorname{Tr}\!\left[ C' \partial^\mu\Pi \right],
    \label{eq:app_L1} \\[4pt]
    \mathcal{L}^{(2)} &= -\frac{i}{2f}\, \partial_\mu\phi \, \operatorname{Tr}\!\left[ C' [\Pi, \partial^\mu\Pi] \right],
    \label{eq:app_L2} \\[4pt]
    \mathcal{L}^{(3)} &= -\frac{\sqrt{2}}{6 f f_\pi}\, \partial_\mu\phi \, \operatorname{Tr}\!\left[ C' [\Pi, [\Pi, \partial^\mu\Pi]] \right].
    \label{eq:app_L3}
\end{align}
The physical roles of the three terms are distinct. The linear piece $\mathcal{L}^{(1)}$ generates derivative kinetic mixing between the ALP and the neutral mesons; it does not correspond to a scattering process and must be removed by field redefinitions before amplitudes are extracted (appendix~\ref{app:diagonalization}). The quadratic piece $\mathcal{L}^{(2)}$ generates three-point $\phi M_i M_j$ vertices and mediates the two-body decays $K \to \pi\phi$. The cubic piece $\mathcal{L}^{(3)}$ generates four-point $\phi M_i M_j M_k$ vertices responsible for the $2\to2$ meson scattering channels and for three-body decays such as $K\to\pi\pi\phi$. All terms carry an overall derivative of $\phi$ and therefore vanish in the soft-ALP limit, as required by the Goldstone nature of $\phi$.

\subsection{Diagonalization of the neutral sector}
\label{app:diagonalization}

\subsubsection{Kinetic mixing}

Evaluating eq.~\eqref{eq:app_L1} with eq.~\eqref{eq:app_meson_matrix} reproduces the kinetic mixing terms from eq.~\eqref{eq:ALP meson interaction},
\begin{align}
    \mathcal{L}^{(1)} ={}& \partial_\mu\phi \left[
      \xi_{\phi\pi^0}\, \partial^\mu\pi^0
    + \xi_{\phi\eta_8}\, \partial^\mu\eta_8
    + \xi_{\phi\eta_0}\, \partial^\mu\eta_0
    + \xi_{\phi K}\left( \partial^\mu K_\phi \right)
    \right],
    \label{eq:app_kinmix}
\end{align}
with the kinetic-mixing coefficients defined in the text as
\begin{align}
    \xi_{\phi\pi^0} &= -\frac{C_{dd}}{2}\frac{f_\pi}{f}, \qquad
    \xi_{\phi\eta_8} = \frac{1}{2\sqrt{3}}\qty(C_{dd} -2\, C_{ss} )\frac{f_\pi}{f}, \notag \\
    \xi_{\phi\eta_0} &= \frac{1}{\sqrt{6}} \qty(C_{dd}+C_{ss})\frac{f_\pi}{f}, \qquad
    \xi_{\phi K} = |C_{sd}|\frac{f_\pi}{f}.
    \label{eq:app_xis}
\end{align}
Since $\phi$ is real while $K^0$ is complex, the ALP mixes only with the single real neutral-kaon direction
\begin{equation}
    K_\phi \equiv \frac{1}{\sqrt{2}|C_{sd}|}\left( C_{sd} K^0 + C_{ds} \bar{K}^0 \right)
    = \sqrt{2}\,\mathrm{Re}\!\left( e^{i\delta_{sd}} K^0 \right),
    \qquad C_{sd} = |C_{sd}| e^{i\delta_{sd}},
    \label{eq:app_Kphi}
\end{equation}
which follows from hermiticity, $C_{ds} = C_{sd}^*$; the orthogonal combination does not mix with $\phi$. For rates that depend only on $|C_{sd}|^2$, such as the thermal reaction densities used below, one may equivalently keep $K^0$ and $\bar K^0$ as complex fields with the Hermitian-conjugate interaction terms always included, which is what we do in appendix~\ref{app:vertices}.

In the $\eta$ sector we define the physical states through,
\begin{equation}
    \eta   = \eta_8 \cos\theta_\eta - \eta_0 \sin\theta_\eta, \qquad
    \eta'  = \eta_8 \sin\theta_\eta + \eta_0 \cos\theta_\eta,
    \label{eq:app_etamix}
\end{equation}
with $\theta_\eta \simeq -19.5^\circ$ ($\sin\theta_\eta \simeq -1/3$) at leading order; our results depend only mildly on the precise value. The mixing coefficients in the physical basis are
\begin{equation}
    \xi_{\phi\eta}  = \xi_{\phi\eta_8}\cos\theta_\eta - \xi_{\phi\eta_0}\sin\theta_\eta,
    \qquad
    \xi_{\phi\eta'} = \xi_{\phi\eta_8}\sin\theta_\eta + \xi_{\phi\eta_0}\cos\theta_\eta .
    \label{eq:app_xi_eta_phys}
\end{equation}

\subsubsection{Field redefinition and physical admixtures}

Consider a single real meson $X \in \{\pi^0, \eta, \eta', K_\phi\}$ with kinetic mixing $\xi \equiv \xi_{\phi X}$,
\begin{equation}
    \mathcal{L} \supset \frac{1}{2}(\partial\phi)^2 + \frac{1}{2}(\partial X)^2
    + \xi\, \partial_\mu\phi\, \partial^\mu X
    - \frac{1}{2} m_\phi^2 \phi^2 - \frac{1}{2} m_X^2 X^2 .
\end{equation}
The unique field redefinition that renders both the kinetic and the mass terms diagonal at $\mathcal{O}(\xi)$ is
\begin{equation}
    X = \hat X + \frac{\xi\, m_\phi^2}{m_X^2 - m_\phi^2}\, \hat\phi,
    \qquad
    \phi = \hat\phi - \frac{\xi\, m_X^2}{m_X^2 - m_\phi^2}\, \hat X,
    \label{eq:app_diag}
\end{equation}
where hatted fields are the canonically normalized mass eigenstates. Two features of eq.~\eqref{eq:app_diag} are important. First, the meson content of the physical ALP,
\begin{equation}
    \theta_{X} = \frac{\xi_{\phi X}\, m_\phi^2}{m_X^2 - m_\phi^2},
    \label{eq:app_theta}
\end{equation}
is suppressed by $m_\phi^2/m_X^2$ and vanishes in the limit $m_\phi \to 0$: a massless derivatively coupled ALP does not inherit unsuppressed meson couplings, as dictated by its Goldstone nature. This is precisely the origin of the $m_\phi^2$ factor in the induced photon coupling $g_{\phi\gamma\gamma}$ of eq.~\eqref{eq:ALPphoton_typeD}, which is obtained by dressing eq.~\eqref{eq:app_theta} with the anomalous $M\gamma\gamma$ vertices of the neutral mesons. Second, the ALP content of the physical mesons, $-\xi_{\phi X}\, m_X^2/(m_X^2-m_\phi^2) \simeq -\xi_{\phi X}$, is \emph{not} suppressed; it is however of order $f_\pi/f$ and only affects observables at $\mathcal{O}(1/f^2)$, beyond the order of this work.

While intermediate quantities such as ``the mixing angle'' depend on the bookkeeping (e.g.\ a symmetric shift followed by a mass-matrix rotation gives the same total admixtures as eq.~\eqref{eq:app_diag}), the combinations in eq.~\eqref{eq:app_diag} are convention independent and are the only ones that enter physical amplitudes. An immediate consequence, used repeatedly below, is that for $m_\phi \ll m_\pi$ the $2\to2$ and $1\to2$ amplitudes are given by the \emph{contact} vertices of $\mathcal{L}^{(2)}$ and $\mathcal{L}^{(3)}$ alone: diagrams in which the ALP attaches to an external meson line through the two-point mixing are proportional to $m_\phi^2$ and negligible, and at leading chiral order there are no three-meson vertices that could generate additional pole topologies.

\subsection{Interaction vertices}
\label{app:vertices}

\subsubsection{Three-point vertices from \texorpdfstring{$\mathcal{L}^{(2)}$}{L2} }

Since $\mathcal{L}^{(2)}$ involves the commutator $[\Pi,\partial_\mu\Pi]$, the singlet $\eta_0$ drops out and only antisymmetric meson pairs appear. The flavor-changing ($\Delta S = 1$) vertices are
\begin{align}
    \mathcal{L}_{K\pi\phi} &= \frac{i C_{sd}}{2f}\, \partial_\mu\phi
    \left[ \left( K^+ \partial^\mu\pi^- - \pi^- \partial^\mu K^+ \right)
    - \frac{1}{\sqrt{2}}\left( K^0 \partial^\mu\pi^0 - \pi^0 \partial^\mu K^0 \right) \right] + \mathrm{h.c.},
    \label{eq:app_Kpiphi} \\[4pt]
    \mathcal{L}_{K\eta_8\phi} &= \frac{i\sqrt{6}\, C_{sd}}{4f}\, \partial_\mu\phi
    \left( K^0 \partial^\mu\eta_8 - \eta_8\, \partial^\mu K^0 \right) + \mathrm{h.c.},
    \label{eq:app_Ketaphi}
\end{align}
mediating $K^+\to\pi^+\phi$, $K^0\to\pi^0\phi$, and, in the physical basis \eqref{eq:app_etamix}, $\eta^{(\prime)}\to K\phi$ (the decays $K\to\eta^{(\prime)}\phi$ are kinematically closed). In addition, $\mathcal{L}^{(2)}$ contains the flavor-diagonal vertices
\begin{align}
    \mathcal{L}^{(2)}_{\rm diag} = \frac{i}{2f}\, \partial_\mu\phi \left[ C_{dd}\left( \pi^+\partial^\mu\pi^- - \pi^-\partial^\mu\pi^+ \right) \right.  &+ C_{ss}\left( K^+\partial^\mu K^- - K^-\partial^\mu K^+ \right) \notag \\
    &+ \left. \left( C_{ss}-C_{dd} \right)\left( K^0\partial^\mu\bar K^0 - \bar K^0\partial^\mu K^0 \right) \right].
    \label{eq:app_L2diag}
\end{align}
Their on-shell matrix elements are proportional to the mass difference of the two mesons and vanish in each degenerate pair; they contribute to physical processes only through off-shell insertions at higher order in $1/f$ and are dropped in the following.

\subsubsection{Four-point vertices from \texorpdfstring{$\mathcal{L}^{(3)}$}{L3}}

The double commutator generates four-point interactions. The pure-pion vertex, controlled entirely by $C_{dd}$ (recall $C_{uu} = 0$ in the type-D model, so the source is not isospin symmetric and a $\phi\pi\pi\pi$ coupling is allowed), reads
\begin{equation}
    \mathcal{L}_{\phi\pi\pi\pi} = \frac{C_{dd}}{3 f f_\pi}\, \partial_\mu\phi
    \left[ 2\pi^+\pi^-\partial^\mu\pi^0 - \pi^0\pi^+\partial^\mu\pi^- - \pi^0\pi^-\partial^\mu\pi^+ \right].
    \label{eq:app_phi3pi}
\end{equation}
This matches the interaction in eq.~\eqref{eq:phi3pifreezein} (up to an overall sign convention). Although the two-pion vertex \eqref{eq:app_L2diag} has no on-shell support for degenerate pions, the four-point coupling \eqref{eq:app_phi3pi} is unsuppressed and drives the dominant mesonic freeze-in channel. $\mathcal{L}^{(3)}$ further contains $\phi\pi K\bar K$ and $\phi\eta_8 K\bar K$ vertices governed by $C_{dd}$ and $C_{ss}$, as well as $\Delta S = 1$ vertices of the type $\phi K\pi\pi$ and $\phi K\eta_8\pi$ proportional to $C_{sd}$. Rather than listing the lengthy vertex Lagrangian, we give in table~\ref{tab:app_amplitudes} the complete set of on-shell $2\to2$ amplitudes obtained from it, which is the information needed to reproduce the freeze-in calculation; the $\Delta S = 1$ four-point vertices are discussed in appendix~\ref{app:neglected}.

\subsection{Amplitudes, cross sections, and reaction densities}
\label{app:amplitudes}

\subsubsection{Two-body decays}

For a generic antisymmetric vertex $\mathcal{L} \supset c\, \partial_\mu\phi\, (A\, \partial^\mu B - B\, \partial^\mu A)$ the amplitude for $A(p_A) \to B(p_B)\, \phi(q)$ is
\begin{equation}
    \mathcal{M} = -i c\; q\cdot(p_A + p_B) = -i c\, (m_A^2 - m_B^2),
    \label{eq:app_decay_amp_general}
\end{equation}
where the $m_\phi$ dependence cancels between $q\cdot p_A$ and $q\cdot p_B$. Applied to eq.~\eqref{eq:app_Kpiphi} this gives
\begin{equation}
    \left|\mathcal{M}(K^+ \to \pi^+\phi)\right| = \frac{|C_{sd}|}{2f}\left( m_K^2 - m_\pi^2 \right),
    \qquad
    \left|\mathcal{M}(K^0 \to \pi^0\phi)\right| = \frac{|C_{sd}|}{2\sqrt{2}f}\left( m_K^2 - m_\pi^2 \right),
    \label{eq:app_Kpi_amp}
\end{equation}
and hence
\begin{align}
    \Gamma(K^+ \to \pi^+\phi) &= \frac{|C_{sd}|^2}{64\pi f^2}\,
    \frac{\lambda^{1/2}(m_K^2, m_\pi^2, m_\phi^2)}{m_K^3}\left( m_K^2 - m_\pi^2 \right)^2 ,
    \label{eq:app_GammaKpi} \\[4pt]
    \Gamma(K^0 \to \pi^0\phi) &= \frac{1}{2}\,\Gamma(K^+ \to \pi^+\phi)\Big|_{m_{K^+},m_{\pi^+}\to m_{K^0},m_{\pi^0}} .
    \label{eq:app_GammaK0pi0}
\end{align}
Eq.~\eqref{eq:app_Kpi_amp} coincides with the standard semileptonic form-factor normalization: writing $\bar s_R\gamma^\mu d_R = \tfrac12\bar s\gamma^\mu d + \tfrac12 \bar s\gamma^\mu\gamma_5 d$, only the vector current contributes between pseudoscalars, and $\langle\pi^+|\bar s\gamma^\mu d|K^+\rangle = F_0^{K\pi}\,(p_K+p_\pi)^\mu$ with $F_0^{K\pi}(0) = 1$ at leading chiral order reproduces eq.~\eqref{eq:app_Kpi_amp} exactly. Using instead the lattice value $F_0^{K\pi}(0) = 0.9698$ changes the rate by $6\%$, which we include in the flavor-constraint analysis of section~\ref{sec:constraints} but neglect in the thermal rates. For the thermal bath, the neutral kaons may equivalently be counted as $(K^0,\bar K^0)$ or $(K_L,K_S)$; the summed production rate depends only on $|C_{sd}|^2$.

The decays $\eta^{(\prime)} \to K^0\phi,\ \bar K^0\phi$ follow from eq.~\eqref{eq:app_Ketaphi} in the same way,
\begin{equation}
    \Gamma(\eta \to K^0\phi) + \Gamma(\eta \to \bar K^0\phi)
    = \frac{3\cos^2\theta_\eta\, |C_{sd}|^2}{64\pi f^2}\,
    \frac{\lambda^{1/2}(m_\eta^2, m_K^2, m_\phi^2)}{m_\eta^3}\left( m_\eta^2 - m_K^2 \right)^2,
    \label{eq:app_Gammaeta}
\end{equation}
with $\cos^2\theta_\eta \to \sin^2\theta_\eta$ and $m_\eta\to m_{\eta'}$ for the $\eta'$. These channels are doubly suppressed---by the small phase space $(m_\eta^2-m_K^2)^3 \ll (m_K^2-m_\pi^2)^3$ or by $\sin^2\theta_\eta$, and by the small thermal abundances of $\eta$ and $\eta'$---and are numerically irrelevant, but we include them for completeness.

\subsubsection{\texorpdfstring{$2\to2$}{2to2} scattering amplitudes}

Table~\ref{tab:app_amplitudes} lists all flavor-conserving $2\to2$ amplitudes generated by $\mathcal{L}^{(3)}$ in the limit $m_\phi \to 0$. For each process $a(p_1)\, b(p_2) \to c(p_3)\, \phi(q)$ we define
\begin{equation}
    s = (p_1+p_2)^2, \qquad t = (p_1-p_3)^2, \qquad u = (p_2-p_3)^2,
    \qquad s+t+u = m_a^2 + m_b^2 + m_c^2 + m_\phi^2 ,
\end{equation}
with the first-listed particle always taken as $a$. Charge-conjugate processes have identical amplitudes up to an irrelevant overall phase. As anticipated, all amplitudes are pure contact terms; they are related to one another by crossing, as can be verified explicitly from the table.

\begin{table}[t]
\centering
\renewcommand{\arraystretch}{1.5}
\begin{tabular}{l l}
\hline\hline
Process $\;a\,b \to c\,\phi$ & Amplitude $\;\mathcal{M} \times \left(4 f f_\pi\right)/i$ \\
\hline
$\pi^+\pi^- \to \pi^0\phi$          & $-2C_{dd}\,(s - m_\pi^2)$ \\
$\pi^\pm\pi^0 \to \pi^\pm\phi$      & $+2C_{dd}\,(m_\pi^2 - t)$ \\
\hline
$\pi^-K^+ \to K^0\phi$,\quad $\pi^+K^- \to \bar K^0\phi$
    & $+\sqrt{2}\left[ C_{dd}\,(m_K^2 - t) + C_{ss}\,(m_\pi^2 - u) \right]$ \\
$\pi^+K^0 \to K^+\phi$,\quad $\pi^-\bar K^0 \to K^-\phi$
    & $+\sqrt{2}\left[ -C_{dd}\,(s - m_K^2) + C_{ss}\,(m_\pi^2 - u) \right]$ \\
$\pi^0K^0 \to K^0\phi$,\quad $\pi^0\bar K^0 \to \bar K^0\phi$
    & $+(C_{dd} - C_{ss})\,(m_\pi^2 - u)$ \\
$\pi^0K^\pm \to K^\pm\phi$
    & $+C_{ss}\,(m_\pi^2 - u)$ \\
\hline
$K^+K^- \to \pi^0\phi$              & $-C_{ss}\,(s - m_\pi^2)$ \\
$K^0\bar K^0 \to \pi^0\phi$         & $-(C_{dd} - C_{ss})\,(s - m_\pi^2)$ \\
$K^0K^- \to \pi^-\phi$,\quad $\bar K^0K^+ \to \pi^+\phi$
    & $+\sqrt{2}\left[ C_{dd}\,(m_K^2 - t) - C_{ss}\,(s - m_\pi^2) \right]$ \\
\hline
$K^+K^- \to \eta_8\phi$             & $-\sqrt{3}\, C_{ss}\,(s - m_\eta^2)$ \\
$K^0\bar K^0 \to \eta_8\phi$        & $+\sqrt{3}\,(C_{dd} - C_{ss})\,(s - m_\eta^2)$ \\
\hline\hline
\end{tabular}
\caption{Flavor-conserving $2\to2$ production amplitudes from $\mathcal{L}^{(3)}$ in the limit $m_\phi\to0$ and in the isospin limit for meson masses. Amplitudes for physical $\eta$ ($\eta'$) final states are obtained from the $\eta_8$ rows by multiplying with $\cos\theta_\eta$ ($\sin\theta_\eta$) and using the physical mass. Charge-conjugate channels carry identical rates. There are no $\pi^0\pi^0$, $\pi^+\pi^+$, or single-$\eta_0$ channels at this order.}
\label{tab:app_amplitudes}
\end{table}

Several features of table~\ref{tab:app_amplitudes} are worth noting. (i) The pure-pion amplitudes are controlled exclusively by $C_{dd}$; for the neutral final state,
\begin{equation}
    \mathcal{M}(\pi^+\pi^- \to \pi^0\phi) = -\frac{i\,C_{dd}}{2 f f_\pi}\left( s - m_\pi^2 \right),
    \label{eq:app_pipi_amp}
\end{equation}
in agreement with eq.~\eqref{eq:pipitopiphi}, and the charged channels follow by crossing $s\to t$. (ii) The neutral-kaon amplitudes involving one pion or one $\eta_8$ are proportional to $C_{dd}-C_{ss}$ and vanish when the source is universal in the $d$--$s$ subspace, $C_{ss}=C_{dd}$; note that because $C_{uu}=0$ this limit is \emph{not} an $SU(3)$-universal source, and accordingly the charged-kaon channels, which are proportional to $C_{ss}$ alone or to independent combinations, survive it. (iii) Since the initial states are distinguishable in every channel (there is no $\pi^0\pi^0$ or $\pi^\pm\pi^\pm$ channel), no identical-particle symmetry factors arise in the collision integrals.

\subsubsection{Cross sections and reaction densities}

For a $2\to2$ process $ab \to c\phi$ with amplitude $\mathcal{M}(s,t)$ the differential cross section is
\begin{equation}
    \frac{d\sigma}{dt} = \frac{|\mathcal{M}(s,t)|^2}{16\pi\, \lambda(s, m_a^2, m_b^2)},
    \label{eq:app_dsigmadt}
\end{equation}
integrated over the kinematic range of $t$ at fixed $s$. For the $t$-independent amplitude \eqref{eq:app_pipi_amp} this yields
\begin{equation}
    \sigma(\pi^+\pi^- \to \pi^0\phi)
    = \frac{C_{dd}^2}{64\pi f^2 f_\pi^2}\,
    \frac{(s - m_\pi^2)^2\, \lambda^{1/2}(s, m_\pi^2, m_\phi^2)}{s\, \lambda^{1/2}(s, m_\pi^2, m_\pi^2)}
    \;\xrightarrow[m_\phi\to0]{}\;
    \frac{C_{dd}^2}{64\pi f^2 f_\pi^2}\,
    \frac{(s - m_\pi^2)^3}{s\, \lambda^{1/2}(s, m_\pi^2, m_\pi^2)},
    \label{eq:app_pipi_xsec}
\end{equation}
reproducing eq.~\eqref{eq:pipitopiphi}; the $t$- and $u$-dependent amplitudes of table~\ref{tab:app_amplitudes} are integrated according to eq.~\eqref{eq:app_dsigmadt} numerically.

The cross sections enter the freeze-in computation through the reaction density of eq.~\eqref{eq:gamma_def},
\begin{equation}
    \gamma(ab \to c\phi) = \frac{T}{64\pi^4} \int_{s_{\rm min}}^\infty ds\, \sqrt{s}\;
    \widehat{\sigma}(s)\, K_1\!\left( \frac{\sqrt{s}}{T} \right),
    \qquad
    \widehat{\sigma}(s) = \frac{2\,\lambda(s, m_a^2, m_b^2)}{s}\, \sigma(s),
    \label{eq:app_reaction_density}
\end{equation}
with $s_{\rm min} = \max[(m_a+m_b)^2, (m_c+m_\phi)^2]$, while decays contribute
\begin{equation}
    \gamma(A \to B\phi) = \frac{g_A\, m_A^2\, T}{2\pi^2}\, K_1\!\left( \frac{m_A}{T} \right) \Gamma(A \to B\phi).
    \label{eq:app_decay_density}
\end{equation}
Our counting conventions are as follows: all mesons are spin-0, $g_A = 1$; every charge eigenstate in table~\ref{tab:app_amplitudes} (and each of $K^\pm$, $K^0$, $\bar K^0$ in the decay channels) is summed as a separate channel with its own equilibrium distribution, so no additional multiplicity factors are applied; particles and antiparticles are counted separately; and Maxwell--Boltzmann statistics are used, consistently with eqs.~\eqref{eq:app_reaction_density}--\eqref{eq:app_decay_density}. We assume kinetic and chemical equilibrium of the hadronic bath, so that all kaon states share a common equilibrium abundance in the isospin limit. The total mesonic yield follows from the sum of eqs.~\eqref{eq:app_reaction_density} and \eqref{eq:app_decay_density} over all channels in the Boltzmann equation of section~\ref{sec:direct_freezein}.

\subsection{Summary of production channels and neglected effects}
\label{app:summary}

\begin{table}[t]
    \centering
    \renewcommand{\arraystretch}{1.35}
    \begin{tabular}{ccc}
    \hline\hline
    Coupling & Vertices & Production channels \\
    \hline
    $C_{dd}, C_{uu}$ & eq.~\eqref{eq:app_phi3pi} & $\pi\pi \to \pi\phi$ \\
    $C_{sd}$ & eqs.~\eqref{eq:app_Kpiphi}, \eqref{eq:app_Ketaphi} & $K^+ \to \pi^+\phi$,\ $K^0 \to \pi^0\phi$,\ $\eta  \to K\phi$ \\
    $C_{dd},\,C_{ss}$ & table~\ref{tab:app_amplitudes} & $\pi K \to K\phi$,\ $K\bar{K} \to \pi\phi$,\ $K\bar K \to \eta \, \phi$ \\
    \hline\hline
\end{tabular}
\caption{Leading mesonic channels entering the freeze-in calculation, organized by the ALP--quark flavor couplings that control them. All channels arise at leading order in the chiral expansion and at $\mathcal{O}(1/f)$.}
\label{tab:app_meson_summary}
\end{table}

Table~\ref{tab:app_meson_summary} organizes the mesonic input by the underlying flavor coupling.
The relative importance of the channels is set by the Wilson coefficients together with the thermal abundances in the hadronic plasma. Pions are the most abundant hadrons throughout the confinement epoch, so the $\pi\pi\to\pi\phi$ channels, controlled solely by $C_{dd}$ and only mildly Boltzmann suppressed, dominate the mesonic contribution; our numerical analysis finds that $\pi^+\pi^-\to\pi^0\phi$ alone contributes approximately \DsharePiPi\ of the final relic abundance. The kaon decays $K\to\pi\phi$, controlled by the flavor-changing coupling $C_{sd}$ characteristic of the Nelson--Barr framework, provide the next-leading contribution, with $\gamma \sim n_K^{\rm eq}\,\Gamma$ suppressed by a single kaon Boltzmann factor. Channels with two initial kaons carry a double Boltzmann suppression $\propto e^{-2m_K/T}$ and are subleading.

\begin{table}[p]
	\centering
	\renewcommand{\arraystretch}{1.25}
	\setlength{\tabcolsep}{6pt}
	\small
	\begin{tabular}{l l c c c c c}
		\hline\hline
		& \textbf{Channel} & $n$ & $Y_\phi$ & $\Omega_\phi h^2$ &
		\textbf{\% of total} & \textbf{$f_\star$ [GeV]} \\
		\hline
		\multirow{12}{*}{Type-U}
		& $\pi^+\pi^-\to\pi^0\phi$              & 1 & $2.31\times10^{-7}$ & $6.37\times10^{-2}$ & $53.10$ & $7.63\times10^{9}$ \\
		& $\pi^0\pi^\pm\to\pi^\pm\phi$          & 2 & $1.46\times10^{-7}$ & $4.04\times10^{-2}$ & $33.66$ & $6.08\times10^{9}$ \\
		& $K^\mp\pi^\pm\to K^0,\bar K^0\phi$    & 2 & $3.01\times10^{-8}$ & $8.30\times10^{-3}$ & $6.92$  & $2.75\times10^{9}$ \\
		& $\bar K^0\pi^-,K^0\pi^+\to K^\mp\phi$ & 2 & $1.32\times10^{-8}$ & $3.64\times10^{-3}$ & $3.03$  & $1.82\times10^{9}$ \\
		& $K^+K^-\to\eta\phi$                   & 1 & $3.51\times10^{-9}$ & $9.68\times10^{-4}$ & $0.81$  & $9.41\times10^{8}$ \\
		& $\bar K^0K^+,K^0K^-\to\pi^\pm\phi$    & 2 & $3.18\times10^{-9}$ & $8.79\times10^{-4}$ & $0.73$  & $8.96\times10^{8}$ \\
		& $K^+K^-\to\pi^0\phi$                  & 1 & $3.01\times10^{-9}$ & $8.32\times10^{-4}$ & $0.69$  & $8.72\times10^{8}$ \\
		& $K^\mp\pi^0\to K^\mp\phi$             & 2 & $2.65\times10^{-9}$ & $7.31\times10^{-4}$ & $0.61$  & $8.17\times10^{8}$ \\
		& $K^\mp\eta\to K^\mp\phi$              & 2 & $1.36\times10^{-9}$ & $3.76\times10^{-4}$ & $0.31$  & $5.86\times10^{8}$ \\
		& $t\to u\phi$                          & 1 & $3.94\times10^{-10}$ & $1.09\times10^{-4}$ & $0.09$ & $3.15\times10^{8}$ \\
		& $c\to u\phi$                          & 1 & $1.77\times10^{-10}$ & $4.88\times10^{-5}$ & $0.04$ & $2.11\times10^{8}$ \\
		& remaining (5 channels)                & 5 & $2.12\times10^{-11}$ & $5.85\times10^{-6}$ & $<0.01$ & $7.31\times10^{7}$ \\
		\hline
		\multirow{16}{*}{Type-D}
		& $\pi^+\pi^-\to\pi^0\phi$                        & 1 & $1.64\times10^{-7}$ & $4.53\times10^{-2}$ & $37.73$ & $7.36\times10^{9}$ \\
		& $\pi^0\pi^\pm\to\pi^\pm\phi$                    & 2 & $1.04\times10^{-7}$ & $2.87\times10^{-2}$ & $23.92$ & $5.86\times10^{9}$ \\
		& $b\to d\phi$                                    & 1 & $8.49\times10^{-8}$ & $2.34\times10^{-2}$ & $19.51$ & $5.30\times10^{9}$ \\
		& $\bar K^0\pi^-,K^0\pi^+\to K^\mp\phi$           & 2 & $2.09\times10^{-8}$ & $5.78\times10^{-3}$ & $4.81$  & $2.63\times10^{9}$ \\
		& $K^\pm\to\pi^\pm\phi$                           & 2 & $2.02\times10^{-8}$ & $5.58\times10^{-3}$ & $4.65$  & $2.59\times10^{9}$ \\
		& $K^0,\bar K^0\to\pi^0\phi$                      & 2 & $1.01\times10^{-8}$ & $2.79\times10^{-3}$ & $2.33$  & $1.83\times10^{9}$ \\
		& $K^\mp\pi^\pm\to K^0,\bar K^0\phi$              & 2 & $9.62\times10^{-9}$ & $2.65\times10^{-3}$ & $2.21$  & $1.78\times10^{9}$ \\
		& $K^\mp\pi\to\pi\phi\ (\Delta S=1)$              & 4 & $3.20\times10^{-9}$ & $8.81\times10^{-4}$ & $0.73$  & $1.03\times10^{9}$ \\
		& $b\to s\phi$                                    & 1 & $2.40\times10^{-9}$ & $6.62\times10^{-4}$ & $0.55$  & $8.90\times10^{8}$ \\
		& $K^0\bar K^0\to\eta\phi$                        & 1 & $2.36\times10^{-9}$ & $6.50\times10^{-4}$ & $0.54$  & $8.82\times10^{8}$ \\
		& $\bar K^0K^+,K^0K^-\to\pi^\pm\phi$              & 2 & $2.03\times10^{-9}$ & $5.59\times10^{-4}$ & $0.47$  & $8.19\times10^{8}$ \\
		& $K^0\bar K^0\to\pi^0\phi$                       & 1 & $2.02\times10^{-9}$ & $5.58\times10^{-4}$ & $0.47$  & $8.18\times10^{8}$ \\
		& $K^0,\bar K^0\pi^0\to K^0,\bar K^0\phi$         & 2 & $1.78\times10^{-9}$ & $4.90\times10^{-4}$ & $0.41$  & $7.67\times10^{8}$ \\
		& $\bar K^0\pi^+,K^0\pi^-\to\pi^\pm\phi$          & 2 & $1.72\times10^{-9}$ & $4.74\times10^{-4}$ & $0.40$  & $7.54\times10^{8}$ \\
		& $K^0,\bar K^0\eta\to K^0,\bar K^0\phi$          & 2 & $9.14\times10^{-10}$ & $2.52\times10^{-4}$ & $0.21$ & $5.50\times10^{8}$ \\
		& remaining (52 channels)                         & 52 & $4.63\times10^{-9}$ & $1.28\times10^{-3}$ & $1.06$ & $1.24\times10^{9}$ \\
		\hline\hline
	\end{tabular}
	\caption{Channel decomposition of the freeze-in abundance at the benchmark point ($m_\phi=1~{\rm MeV}$, $T_{\rm sw}=155~{\rm MeV}$, $\sqrt{s}<\Lambda_\chi$ in the hadronic channels), for the type-U (upper) and type-D (lower) models. The column $n$ gives the number of charge states summed in each row; $Y_\phi$, $\Omega_\phi h^2$ and the percentage refer to that sum. Yields are quoted at the decay constant obtained including all channels, $f=1.05\times10^{10}$~GeV (type-U) and $1.20\times10^{10}$~GeV (type-D); $f_\star$ is the decay constant at which the row alone would reproduce the observed abundance, $f_\star=f\sqrt{Y_{\rm row}/Y_{\rm tot}}$. The enumeration is complete: every four-point vertex generated by $\mathcal{L}^{(3)}$ contributes three crossings, giving 75 kinematically open scattering channels in type-D and 16 in type-U, plus 16 channels with an $\eta'$ in the initial state that are closed by $\sqrt{s}<\Lambda_\chi=4\pi f_\pi$. Rows below $0.2\%$ are collected in ``remaining''; the full list is provided as ancillary material. Type-U has no mesonic decay channels, since $C^{(U)}$ is diagonal in the light-quark sector and generates no $\Delta S=1$ vertex.}
	\label{tab:channels}
\end{table}

\paragraph{Neglected contributions.}
\label{app:neglected}
The following effects arise at the same order in the chiral expansion but are quantitatively negligible:
\begin{itemize}
\item \emph{Three-body decays.} The $\Delta S=1$ four-point vertices in $\mathcal{L}^{(3)}$ generate $K \to \pi\pi\phi$ at the same order as the $2\to2$ scatterings. Integrating the corresponding Dalitz distributions we find, for $m_\phi \ll m_\pi$,
\begin{equation}
    \frac{\Gamma(K^0\to\pi^+\pi^-\phi)}{\Gamma(K^0\to\pi^0\phi)} \simeq 1.7\times10^{-3},
    \qquad
    \frac{\Gamma(K^0\to\pi^0\pi^0\phi)}{\Gamma(K^0\to\pi^0\phi)} \simeq 0.9\times10^{-3},
\end{equation}
so that three-body decays modify the kaon-decay reaction density at the per-mille level; $\eta\to\pi\pi\phi$ is suppressed both by phase space and by $n_\eta \ll n_K$.
\item \emph{$\Delta S = 1$ scatterings.} The same vertices induce scatterings such as $\pi K \to \pi\phi$ and $K\bar K \to K\phi$, proportional to $|C_{sd}|^2$ like the kaon decays. Their reaction densities are suppressed relative to the decay density by $\sim n_\pi \langle\sigma v\rangle / \Gamma$, which we estimate to be at or below the percent level for $T \lesssim m_K$; we therefore neglect them.
\item \emph{Higher-order corrections.} Chiral loops are suppressed by $p^2/(4\pi f_\pi)^2$ and introduce no new channels at $\mathcal{O}(1/f)$. Isospin-violating $\pi^0$--$\eta$ mixing, of order $(m_d-m_u)/(m_s-\hat m)$, shifts individual amplitudes at the percent level. Effects of order $m_\phi^2/m_\pi^2$, including the mixing-induced contributions of eq.~\eqref{eq:app_theta} to scattering amplitudes, are negligible for the freeze-in mass range $m_\phi \ll m_\pi$.
\end{itemize}
All of these lie below the precision target of the present analysis. Together with the partonic channels of section~\ref{sec:direct_freezein}, the decay widths \eqref{eq:app_GammaKpi}--\eqref{eq:app_Gammaeta} and the amplitudes of table~\ref{tab:app_amplitudes} constitute the full leading-order input to the freeze-in Boltzmann equation used in this work.

\subsection{Remark on the type-U model}
\label{app:typeU_remark}

For the type-U model the matching proceeds identically with $C' \to C'^{(U)} = \operatorname{diag}(C^{(U)}_{uu}, 0, 0)$, cf.~eq.~\eqref{eq:ALP quark coupling type-U} and the definition of $C'^{(U)}$ in the text. All flavor-changing meson vertices then vanish, so there are no mesonic decay channels. However, the commutator structure does \emph{not} eliminate the four-point interactions: eq.~\eqref{eq:app_L3} yields
\begin{equation}
    \mathcal{L}^{(U)}_{\phi\pi\pi\pi} = -\frac{C^{(U)}_{uu}}{3 f f_\pi}\, \partial_\mu\phi
    \left[ 2\pi^+\pi^-\partial^\mu\pi^0 - \pi^0\pi^+\partial^\mu\pi^- - \pi^0\pi^-\partial^\mu\pi^+ \right],
    \label{eq:app_phi3pi_U}
\end{equation}
i.e.\ the same vertex as eq.~\eqref{eq:app_phi3pi} with $C_{dd} \to -C^{(U)}_{uu}$ (for a general source the coefficient is proportional to $C_{dd}-C_{uu}$), together with the corresponding $\pi K$ and $K\bar K$ channels with $C_{dd}\to C_{uu}$, $C_{ss}\to 0$ in table~\ref{tab:app_amplitudes}. Since $|U^{(U)}_{R,41}| \simeq 1$ implies $C^{(U)}_{uu} = \mathcal{O}(1)$, the $\pi\pi \to \pi\phi$ channel contributes to freeze-in production in the type-U model as well.

\section{Freeze-in production: channel-by-channel breakdown}
\label{app:freezein_channels}

For completeness, we record here the full channel decomposition of the freeze-in relic abundance at the two benchmark points, together with the yield-evolution histories from which the tables of section~\ref{sec:direct_freezein} are obtained. All quantities are computed at $m_\phi=1~{\rm MeV}$ with the switching temperature $T_{\rm sw}=155~{\rm MeV}$ and the chiral cutoff $\Lambda_\chi=4\pi f_\pi$; quark channels are integrated for $T>T_{\rm sw}$ and mesonic channels for $T<T_{\rm sw}$ (section~\ref{sec:direct_freezein}). For each channel we quote the yield $Y_\phi$, the induced $\Omega_\phi h^2$, its percentage of the total, and the decay constant $f_\star$ at which that channel alone would reproduce the observed abundance ($f_\star\propto\sqrt{Y_\phi}$, so a large $f_\star$ marks an efficient channel).

\bibliography{ref}
\bibliographystyle{JHEP}
\end{document}